\documentclass[12pt,preprint]{aastex}
\usepackage{epsfig,amssymb}
\usepackage{graphicx}
\usepackage{rotating}
\newcommand{\ben}{\begin{enumerate}}
\newcommand{\een}{\end{enumerate}}
\newcommand{\bfig}{\begin{figure}}
\newcommand{\efig}{\end{figure}}
\newcommand{\beq}{\begin{equation}}
\newcommand{\eeq}{\end{equation}}
\newcommand{\mbf}{\mathbf}

\newcommand{\cs}{\mathcal{S}}

\newcommand{\cv}{\mathcal{V}}
\renewcommand{\thefootnote}{\dag}

\shorttitle{Active-Region Energetics and Helicity I}
\shortauthors{Georgoulis \& LaBonte}

\begin{document}
\title{Magnetic Energy and Helicity Budgets in the Active-Region Solar Corona. I. 
Linear Force-Free Approximation}


\author{Manolis K. Georgoulis \& Barry J. LaBonte\footnote{Deceased October 24, 2005}}
\affil{The Johns Hopkins University Applied Physics Laboratory,\\
11100 Johns Hopkins Rd. Laurel, MD 20723-6099, USA}
\begin{abstract}
We self-consistently derive the magnetic energy and relative magnetic
helicity budgets of a three-dimensional linear force-free magnetic
structure rooted in a lower boundary plane. For the potential magnetic
energy we derive a general expression that gives results practically  
equivalent to those of the magnetic Virial theorem. All magnetic 
energy and helicity budgets are
formulated in terms of surface integrals applied to 
the lower boundary, thus avoiding computationally intensive
three-dimensional magnetic field extrapolations. 
We analytically and numerically connect our derivations with
classical expressions for the magnetic energy and helicity, thus
presenting a so-far lacking unified treatment of the energy/helicity 
budgets in the constant-alpha approximation. Applying our
derivations to photospheric vector magnetograms of an eruptive and a
noneruptive solar active regions, we find that the most profound
quantitative difference between these regions lies in the estimated
free magnetic energy and relative magnetic helicity budgets.
If this result is verified with a large number of active regions, it
will advance our understanding of solar eruptive phenomena. 
We also find that the constant-alpha approximation gives rise 
to large uncertainties in the calculation of the free 
magnetic energy and the relative magnetic helicity. 
Therefore, care must be exercised when this approximation is 
applied to photospheric magnetic field observations. 
Despite its shortcomings, the constant-alpha approximation is adopted
here because this study will form the basis of a comprehensive 
nonlinear force-free description of the energetics and helicity 
in the active-region solar corona, which is our ultimate objective. 
\end{abstract}
\section{Introduction}
\renewcommand{\thefootnote}{\arabic{footnote}}
\setcounter{footnote}{0}
The magnetic origin of solar eruptions has been
established over the past several decades of solar research. 
Most eruptions originate in  
active regions that are, in general, closed magnetic structures 
rooted in the solar photosphere. Magnetized plasma motions in
the solar atmosphere prevent a magnetic structure from attaining a 
minimum-energy, current-free state. 
Excess magnetic energy in active regions 
is manifested by the appearance of electric currents (Leka et al. 1996).
Eruptive and noneruptive manifestations, such as coronal mass ejections
(CMEs) and confined solar flares, respectively, 
must be fueled from this reservoir of free magnetic energy that is 
thought to be released in intermittent episodes of magnetic reconnection.

A popular, early, view of a magnetic energy release event 
involved a nonpotential pre-event state relaxing into 
a potential, or nearly potential, post-event state of the magnetic 
configuration. The excess (nonpotential) magnetic energy was thought
to be released during the relaxation. Seminal works on magnetic helicity 
(Woltjer 1958; C\v{a}lug\v{a}reanu 1961; Berger \& Field 1984; Finn \& 
Antonsen 1985; Berger 1985; 1988; Moffatt \& Ricca 1992 and others), 
however, demonstrated 
that this view is incomplete or even misleading: magnetic helicity relates to the 
linkage of a magnetic structure (twist, torsion, and 
writhe), which is globally {\it invariant} even under resistive processes, 
such as magnetic reconnection. Helicity is present wherever electric
currents are present. Therefore, an isolated helical magnetic structure cannot 
relax to a potential state unless its magnetic helicity is bodily removed 
from it. This provides a plausible interpretation for CMEs (Low 1994; 
Rust 1994a; 1994b), provided that the magnetic helicity in the
erupting structures is not transferred to other parts of the solar
atmosphere along preexisting or reconnected magnetic field lines. 
Indeed, eruptive activity seems to be necessary
for the Sun where differential rotation and subsurface dynamo
continuously generate helicity in the two solar hemispheres 
(Berger \& Ruzmaikin 2000), with a statistical
hemispheric segregation exhibited by magnetic structures of opposite
senses of helicity (Pevtsov, Canfield, \&
Metcalf 1995). In confined events the magnetic 
configuration cannot relax to the potential state but, at best, it may relax 
to the lowest possible energy state that preserves the pre-event amount of magnetic 
helicity. This is known to be a constant-alpha, 
linear force-free (LFF) state (Woltjer 1958; Taylor 1974; 1986). Several  
works rely on the Woltjer-Taylor theorem, although controversy 
remains over its applicability 
to the Sun (Kusano et al. 1994, but also Antiochos \& DeVore 1999). 

Whether magnetic helicity {\it per ce} is important for 
solar eruptions is also a subject of debate 
(e.g. Rust 2003; Rust \& LaBonte 2005, but also Phillips, MacNeice, \& 
Antiochos 2005). Regardless, however, knowledge of the 
magnetic helicity is essential for a complete assessment of the
magnetic complexity present in the solar atmosphere. 
Berger \& Field (1984) and Finn \& Antonsen (1985) derived a
gauge-invariant definition of magnetic helicity applying to open and
multiply connected volumes such as the ones assumed for the solar
atmosphere. The resulting {\it relative} magnetic helicity subtracts the 
helicity of the reference (potential) field so a nonzero value 
implies by definition the presence of free magnetic energy in the configuration. 
The relative magnetic helicity $H_m$ has two equivalent forms in
the above works, namely 
\beq
H_m = \int _{\cv} (\mbf{A} \pm \mbf{A_p}) \cdot (\mbf{B} \mp \mbf{B_p}) d \cv\;\;,
\label{in1}
\eeq
where $\mbf{B_p}$ and $\mbf{A_p}$ are the potential magnetic field and its 
generating vector potential, respectively, and $\mbf{B}$, $\mbf{A}$ are the 
respective quantities of the nonpotential field. The integration refers to the 
open volume $\cv$ that contains the part of the magnetic structure 
extending above a lower boundary. To derive equation (\ref{in1}), 
both Berger \& Field (1984) and Finn \& Antonsen (1985) assumed nonlinear 
force-free (NLFF) magnetic fields. The force-free approximation is
probably necessary for helicity calculations because only in this
case one obtains some knowledge of the magnetic field vector and the 
generating vector potential required to evaluate equation (\ref{in1}). 

Equation (\ref{in1}) cannot be evaluated in the active-region
atmosphere, however, because the magnetic field vector is unknown 
above the lower boundary, be it the photosphere or the low 
chromosphere. 
Currently, active-region magnetic fields can only be measured in 
this boundary, so the only way to evaluate the relative magnetic helicity through 
equation (\ref{in1}) is by force-free (preferably NLFF) 
field extrapolation into the active-region corona 
using the measured magnetic fields as the required boundary
condition. However, the NLFF extrapolation of observed solar magnetic 
fields remains an active research area where even the most successful
of the existing techniques (Schrijver et al. 2006 and references
therein) are too slow to fully exploit the spatial resolution of
modern (let alone, future) magnetographs. The NLFF approximation should 
always be pursued given that the LFF approximation is almost certainly an 
oversimplification for most active-region fields and it can even be misleading in 
several cases. Even the NLFF approximation is most likely invalid in the 
photosphere (Georgoulis \& LaBonte 2004), although it may hold in and 
above the chromosphere (Metcalf et al. [1995]; see, however, Socas-Navarro [2005]). 

Even in case equation (\ref{in1}) is evaluated, however, 
it does not establish a  
link between the relative magnetic helicity and the magnetic free energy of the 
studied configuration. In addition, it might be risky 
to evaluate a volume integral 
of extrapolated fields at large heights above the boundary because 
numerical effects might settle in and affect the result. 
One, therefore, envisions 
a convenient {\it surface-integral} representation of the relative 
magnetic helicity that might alleviate the need for full-fledged 
three-dimensional extrapolations. To our 
knowledge, this has been attempted only in the LFF 
approximation following either the theoretical analysis of 
Berger (1985) or the ``twist'' helicity of Moffatt \& Ricca
(1992). In the first case (D\'{e}moulin et al. 2002; Green et
al. 2002), the employed formula for the relative magnetic helicity is 
\beq
H_m= 2 \alpha \sum _{l=1}^{n_x} \sum _{m=1}^{n_y} 
{{|b^2_{u_l,v_m}|} \over {(u_l^2 + v_m^2)^{3/2}}}\;\;,
\label{in2}
\eeq
where $\alpha$ is the unique, representative value of the force-free
parameter and 
$b_{u_l,v_m}$ is the Fourier amplitude of the measured normal magnetic 
field for the harmonic $(u_l,v_m)$ in a two-dimensional Fourier space
with linear dimensions $n_x$, $n_y$. Equation (\ref{in2}) is a
linearized version of the actual formula of Berger
(1985). Linearization helps avoid $|H_m| \rightarrow \infty$ when 
$|\alpha| \rightarrow (2 \pi /L)$, where $L$ is the linear size of
the computational domain. Detailed discussions and an extension of
Berger's (1985) analysis will be given in \S4.2 and Appendix B. 
In the second case (R\'{e}gnier, Amari, \& Canfield 2005) the 
magnetic helicity is approximated by the ``twist'' helicity
of a semi-circular, constant-alpha magnetic flux tube, i.e., 
\beq
H_m = {{1} \over {8}} \alpha L \Phi ^2\;\;,
\label{in3}
\eeq
where $L$ is the characteristic footpoint separation length of the
tube and $\Phi$ is the magnetic flux carried by the tube. Though
useful, equations (\ref{in2}) and (\ref{in3}) also lack a much wanted link
between the relative magnetic helicity and the free magnetic energy in
the LFF magnetic structure that would enable a complete, 
self-consistent, description of its energetics. 
Moreover, it is not clear how to 
generalize equations (\ref{in2}) and (\ref{in3}) into a 
NLFF calculation that, as should be always
kept in mind, must be the ultimate objective of the calculation. 

The above difficulties and lack of information in the calculation of the 
{\it total} relative magnetic helicity prompted 
alternative lines of research. The lower boundary of a closed 
magnetic structure, where all magnetic field lines are supposed to be rooted, acts 
as the driver of the evolution in the structure either via boundary
flows or via the injection of additional structure through it. 
Therefore, magnetic helicity can 
either be transported to and from the structure through this boundary or it can be  
generated by flows on the boundary{\footnote{In the Sun, generation of 
helicity above the photosphere automatically 
implies the generation of an equal and opposite amount of helicity 
below the photosphere, 
to ensure a zero net helicity.}}. Based on these principles, 
Berger \& Field (1984) derived a surface-integral expression for the 
temporal variation $(dH_m/dt)$ of the relative magnetic helicity 
in a magnetic configuration. Besides its dependence on magnetic 
field vectors and vector potentials, as in equation (\ref{in1}), 
$(dH_m/dt)$ depends on the boundary flows. The advantage of 
the Berger \& Field (1984) expression 
for $(dH_m/dt)$ is that it does not explicitly require force-free fields.  
The calculation of $(dH_m/dt)$ has been attempted by numerous 
authors over the past few years 
(see, e.g., Nindos 2006; 
LaBonte, Georgoulis, \& Rust 2007, and references therein), 
although it suffers from the lack of a reference value, namely the 
{\it total} relative magnetic helicity. 
The total relative helicity is a focus of this work, 
so the formula of Berger \& Field (1984) will not be discussed
further. 

This study is the first of a series of studies 
that perform a self-consistent 
calculation of the total magnetic energy and relative magnetic helicity in 
a closed magnetic configuration. 
We devise a practical way to calculate magnetic 
energies and helicities from solar active-region vector magnetograms 
provided that the observed magnetic configuration 
is isolated and flux-balanced on the presumed ``plane'' of the observations.  
The final expressions for the magnetic energy and helicity are 
derived in the form of surface, rather than volume, integrals. 
We always assume that the studied magnetic field configuration is in a
force-free equilibrium. In this study, we provide the analytical
foundation of a constant-alpha, LFF, energy-helicity calculation. 
A NLFF generalization of the energy-helicity equations 
will be the subject of a later study. The LFF analysis of this work 
relies on the energy-helicity formula of Berger (1988) evaluated via an
application and extension of Berger's (1985) 
analysis for the magnetic energies and
the relative magnetic helicity. 
Our objective in this work is 
the derivation of practical LFF energy and helicity equations 
that can be readily applied to solar vector magnetogram data.
The magnetic energy budgets for a constant-alpha 
magnetic configuration are discussed in \S2. 
The LFF energy-helicity formula is discussed in
\S3. The relative magnetic helicity is derived both as a volume and as
a surface integral in \S4. In \S5 we apply our LFF analysis to 
vector magnetograms of two solar active regions 
and in \S6 we summarize and discuss our analysis and results. 
\section{Gauge-invariant definitions and the magnetic energy equation}
\subsection{Magnetic field and the vector potential}
Assuming planar geometry, consider a magnetic structure $\mbf{B}$
extending in the half space $z \ge 0$ above a lower boundary $\cs$ ($z=0$). 
Let an open volume $\cv$ of the half space $z \ge 0$ include the
structure and extend to infinity with its only boundary being the
surface $\cs$. If $\cs$ is not a flux (magnetic) surface, i.e. if
$\mbf{B} \cdot \mbf{\hat{z}}|_{\cs} \ne 0$, then the configuration 
is analogous to a 
solar magnetic structure rooted in a small (assumed planar) part $\cs$
of the photosphere and extending to infinity above $\cs$. Here 
$\mbf{\hat{z}}$ is the unit vector along the $z$-axis of a 
Cartesian coordinate system with an arbitrary origin on $\cs$. 
In the absence of plasma, this magnetic
configuration can only be the vacuum, current-free magnetic field
$\mbf{B_p}$ if the configuration is isolated (not interacting with
other configurations) and flux-balanced on
$\cs$. The presence of plasma dictates a current-carrying
magnetic structure $\mbf{B_c}$, such that 
\beq
\mbf{B}=\mbf{B_p}+\mbf{B_c}\;\;.
\label{Bdef}
\eeq
The divergence-free properties of $\mbf{B}$ and $\mbf{B_p}$ together
with equation (\ref{Bdef}) ensure that $\mbf{B_c}$ is also
divergence-free. As a result, we can define generating vector
potentials $\mbf{A_p}$, $\mbf{A}$, and $\mbf{A_c}$ for 
$\mbf{B_p}$, $\mbf{B}$, and $\mbf{B_c}$, respectively. 
By definition, 
\begin{mathletters}
\beq
\nabla \times \mbf{A_p}=\mbf{B_p} \;\;,
\eeq
\beq
\nabla \times \mbf{A}=\mbf{B}\;\;.
\eeq
\end{mathletters}
%
In addition, the Coulomb gauge is adopted for both $\mbf{A_p}$ and
$\mbf{A}$ to provide 
\begin{mathletters}
\beq
\nabla \cdot \mbf{A_p}=0\;\;,
\eeq
\beq
\nabla \cdot \mbf{A}=0\;\;.
\eeq
\end{mathletters}
%
Since $\cs$ is
not a flux surface, however, both the definition of $\mbf{B_p}$ and
some topological properties of the field, most notably the ones
present in its magnetic helicity integral, are not unique
(gauge-invariant) and hence lack a physical meaning 
(e.g. Dixon et al. 1989; Berger 1999). Berger (1988) addressed the
problem by providing gauge conditions for $\mbf{A}$ and $\mbf{A_p}$
such that both $\mbf{B_p}$ and the magnetic helicity can be uniquely
defined. These conditions are 
$\mbf{A_p} \cdot \mbf{\hat{n}}|_{\partial \cv} =0$ and 
$\mbf{A_p} \times \mbf{\hat{n}}|_{\partial \cv} = 
\mbf{A} \times \mbf{\hat{n}}|_{\partial \cv}$ and were formulated for
a volume  $\cv$ bounded by a surface $\partial \cv$, where 
$\mbf{\hat{n}}$ is the unit vector normal to $\partial \cv$ and
oriented outward from $\cv$. If $\cv$ extends
to infinity, Berger (1988) stresses that $\mbf{A_p}$ and $\mbf{A}$
{\it must} additionally vanish at infinity. This
restricts the above gauge conditions to the lower boundary $\cs$, 
so that 
\begin{mathletters}
\beq
\mbf{A_p} \cdot \mbf{\hat{z}}|_{\cs}=0\;\;,
\eeq
\beq
\mbf{A_p} \times \mbf{\hat{z}}|_{\cs}=\mbf{A} \times
\mbf{\hat{z}}|_{\cs}\;\;,
\eeq
\end{mathletters}
%
where $\mbf{\hat{z}}=-\mbf{\hat{n}}$. From equation (\ref{Bdef}) and
the conditions of equations (5)-(7) we can derive additional
conditions for the vector potential $\mbf{A_c}$. Writing equation
(\ref{Bdef}) in terms of vector potentials, we obtain 
$\mbf{A} = \mbf{A_p} + \mbf{A_c} + \nabla \phi$, where $\phi$ is an
arbitrary scalar. Choosing the gauge such that $\nabla \phi =0$, one
obtains 
\beq
\mbf{A} = \mbf{A_p} + \mbf{A_c}\;\;,
\label{bcon4}
\eeq 
If one now takes the dot and cross products of equation (\ref{bcon4})
with $\mbf{\hat{z}}$, then one obtains
\begin{mathletters}
\beq
\mbf{A_c} \cdot \mbf{\hat{z}}|_{\cs}=\mbf{A} \cdot \mbf{\hat{z}}|_{\cs}\;\;,
\eeq
\beq
\mbf{A_c} \times \mbf{\hat{z}}|_{\cs}=0\;\;,
\eeq
\end{mathletters}
%
where we have used equations (7a) and (7b) to reach equations (9a) and
(9b), respectively. 

For a volume $\cv$ bounded by a surface $\partial \cv$, another
condition for the uniqueness (gauge-invariance) of $\mbf{B_p}$ and the 
magnetic helicity is that $\mbf{B}$ and $\mbf{B_p}$ share the same
normal component on $\partial V$, i.e. 
$\mbf{B} \cdot \mbf{\hat{n}}|_{\partial \cv} = 
 \mbf{B_p} \cdot \mbf{\hat{n}}|_{\partial \cv}$. In our case, where
$\cv$ is only bounded by $\cs$ and extends to infinity above $\cs$ ($z >0$), 
and for $\mbf{A_p}$, $\mbf{A}$, $\mbf{B_p}$, $\mbf{B}$ vanishing at
infinity, the condition refers only to the boundary $\cs$, i.e. 
\beq
\mbf{B} \cdot \mbf{\hat{z}}|_{\cs}=\mbf{B_p} \cdot \mbf{\hat{z}}|_{\cs}\;\;. 
\label{bcon2}
\eeq
It is not necessary to independently pose equation (\ref{bcon2}),
however, as it stems directly from equations (5) and (7b). Take the
normal (vertical) components of equation (5a) and (5b) on $\cs$ to
obtain 
\beq
\nabla _h \cdot (\mbf{A_p} \times \mbf{\hat{z}})|_{\cs} = \mbf{B_p} \cdot \mbf{\hat{z}}|_{\cs}
\;\;\;\;,\;\;\;\;
\nabla _h \cdot (\mbf{A} \times \mbf{\hat{z}})|_{\cs} = \mbf{B} \cdot \mbf{\hat{z}}|_{\cs}\;\;,
\label{bcon3}
\eeq
respectively, where $\nabla _h$ denotes differentiation on the 
horizontal plane, i.e. the boundary $\cs$. Equation
(\ref{bcon2}) immediately follows from combining equations (7b) and
(\ref{bcon3}). In addition, from equations (\ref{Bdef}) and (\ref{bcon2})
we obtain $\mbf{B_c} \cdot \mbf{\hat{z}}|_{\cs}=0$, so
$\mbf{B_c}$ is a {\it closed}, {\bf toroidal} magnetic field 
on $\cs$. In fact, $\mbf{B_c}$ is purely toroidal in any cross-section
$\cs '$ of $\cv$. This has been concluded by Kusano et al. (1994) and
Berger (1999) who argued that the net toroidal flux of
$\mbf{B_p}$ and $\mbf{B}$ should be the same along any cross-section
of $\cv$. The potential field $\mbf{B_p}$ being purely {\bf poloidal}, on
the other hand, one expects $\mbf{B_p} \cdot \mbf{B_c}=0$. 
A construction of $\mbf{B_p}$ and $\mbf{B_c}$ by poloidal and toroidal
components, respectively, can also be found in Berger (1985).
\subsection{Magnetic energy and helicity budgets}
From the general equation (\ref{Bdef}) it is clear that 
the total 
magnetic energy $E=[1/(8 \pi)] \int _{\cv} B^2 d \cv$ of a 
closed magnetic structure is simply the sum of the potential magnetic energy 
$E_p=[1/(8 \pi)] \int _{\cv} B_p^2 d \cv$ and the nonpotential magnetic energy 
$E_c=[1/(8 \pi)] \int _{\cv} B_c^2 d \cv$ 
stored in the configuration in the form of electric currents:
\beq
E=E_p+E_c\;\;.
\label{eneq}
\eeq
Our objective will be to derive a convenient expression for each of the terms 
in equation (\ref{eneq}). In this section we provide general energy
expressions enabled by the 
gauge invariant definitions of the vector potentials $\mbf{A}$, 
$\mbf{A_p}$, and $\mbf{A_c}$. Equations for the 
potential energy $E_p$ can be directly
applied to solar magnetic field measurements. Applicable expressions for
the total energy $E$ of the magnetic structure are given in the
following sections, where the LFF approximation is adopted. 

From the definition of $A_p$, equation (5a), the potential magnetic
energy of the configuration is given by 
\beq
E_p = {{1} \over {8 \pi}} \int _{\partial \cv} 
\mbf{A_p} \times \mbf{B_p} \cdot \mbf{\hat{n}}\;d \sigma\;\;,
\label{pen1a}
\eeq
where $d \sigma$ is the surface element on $\partial \cv$. To reach
equation (\ref{pen1a}) we have used Gauss's theorem and the
current-free condition, $\nabla \times \mbf{B_p}=0$.
Since $\mbf{A_p}$ and $\mbf{B_p}$, together with $\mbf{A}$ and
$\mbf{B}$, all vanish at infinity, however, 
the above surface integral applies
only to the lower boundary $\cs$. The potential energy $E_p$ becomes,
therefore, 
\beq
E_p={{1} \over {8 \pi}}
\int _{\cs} \mbf{B_p} \times \mbf{A_p} \cdot \mbf{\hat{z}} \; d \cs\;\;. 
\label{pen1}
\eeq

Similarly, from the definition of $\mbf{A}$, equation (5b), the total
magnetic energy of the configuration is given by (see also Berger 1988) 
\beq
E={{1} \over {8 \pi}} \int _{\cs} \mbf{B} \times \mbf{A_p} \cdot \mbf{\hat{z}} \;d \cs 
+ {{1} \over {8 \pi}} \int _{\cv} \mbf{A} \cdot \nabla \times \mbf{B}\;d \cv\;\;. 
\label{ten2}
\eeq
The total energy from equation (\ref{ten2}) naturally tends to  
the potential energy in case the magnetic field vector $\mbf{B}$ tends
to its current-free limit $\mbf{B_p}$. Decomposing $\mbf{B}$ into
$\mbf{B_p}$ and $\mbf{B_c}$, one may derive equation (\ref{eneq}), 
where the potential energy is given by equation (\ref{pen1}) and 
$E_c =[1/(8 \pi)] \int _{\cv} B_c^2 d \cv$.

In view of the definitions and conditions for $\mbf{A}$ and
$\mbf{A_p}$, equations (5)-(7), on the other hand, the relative
magnetic helicity of equation (\ref{in1}) can be written as (e.g.,
Berger 1999)  
\beq
H_m= \int _{\cv} \mbf{A} \cdot \mbf{B} d \cv\;\;, 
\label{hm}
\eeq
 
Equation (\ref{pen1}) enables the calculation of the
potential energy $E_p$ for a flux-balanced magnetic 
configuration $\mbf{B}$, regardless of whether the magnetic
field vector $\mbf{B}$ is fully known on $\cs$. What is needed 
is the boundary condition for the vertical field 
$B_z=\mbf{B} \cdot \mbf{\hat{z}}$ on $\cs$, that uniquely determines 
the potential magnetic field $\mbf{B_p}$ and its vector potential 
$\mbf{A_p}$ on the boundary. In particular, assuming 
$\mbf{B_p} = -\nabla \psi$, where $\psi$ is a smooth scalar, 
$\nabla ^2 \psi =0$, Schmidt (1964) showed that 
\beq
\psi (\mbf{r},z) = {{1} \over {2 \pi}} \int \int 
{{B_z (\mbf{r'}) dx'dy'} \over \sqrt{(\mbf{r} - \mbf{r'})^2+ z^2}}\;\;,
\label{sc1}
\eeq
where $\mbf{r}=x \mbf{\hat{x}} + y \mbf{\hat{y}}$, 
$\mbf{r'}=x' \mbf{\hat{x}} + y' \mbf{\hat{y}}$ are vector positions on
$\cs$, defined for a given Cartesian coordinate system centered on
$\cs$. For the vector potential $\mbf{A_p}$ one similarly obtains (see
also DeVore 2000)
\beq
\mbf{A_p}(\mbf{r},z)= \nabla \times \mbf{\hat{z}} 
\int _z ^{\infty} \psi (\mbf{r},z') dz'\;\;.
\label{sc2}
\eeq
Although exact, equations (\ref{sc1}) and (\ref{sc2}) are
computationally extensive. Much faster alternatives are provided by
means of Fourier transforms. Alissandrakis (1981), in particular,
showed that 
\beq
\mbf{B_p}(\mbf{r})
          = F^{-1} ({{-iu} \over {\sqrt{u^2+v^2}}} b_{u,v}) \mbf{\hat{x}} + 
            F^{-1} ({{-iv} \over {\sqrt{u^2+v^2}}} b_{u,v}) \mbf{\hat{y}} +
            B_z \mbf{\hat{z}}\;\;,
\label{al81}
\eeq
while Chae (2001) showed that 
\beq
\mbf{A_p}(\mbf{r})= F^{-1}({{iv} \over {u^2+v^2}} b_{u,v}) \mbf{\hat{x}} + 
                          F^{-1}({{-iu} \over {u^2+v^2}} b_{u,v})\mbf{\hat{y}}\;\;,
\label{ch01}
\eeq
where $b_{u,v}= \sum  _{l=1}^{L_x} \sum _{m=1}^{L_y} B_{z_{m,l}}
exp[-i(ul+vm)]$ is the Fourier amplitude of $B_z$, 
$u=(2 \pi l/L_x)$, $v=(2 \pi m/L_y)$, $L_x$, $L_y$
are the linear dimensions of $\cs$, 
and $F^{-1}(g)$ denotes the inverse Fourier
transform of a function $g$. Albeit much faster, however, equations
(\ref{al81}) and (\ref{ch01}) assume periodic boundary conditions for
$\mbf{B_p}$ and $\mbf{A_p}$ which contradicts the
assumption of $\mbf{A}$ and $\mbf{B}$ vanishing at
infinity. This problem is also well known. To mitigate the effects of
the periodic boundary conditions assumed when Fourier transforms are
used, one typically
surrounds the initial flux concentration with a region of zero
flux. In our calculations in \S5.2 we have applied equations
(\ref{al81}) and (\ref{ch01}) to zero-buffered magnetograms of solar 
active regions.   

From the above, $E_p$ can be readily calculated for 
flux-balanced photospheric or chromospheric 
(not necessarily vector) magnetograms of solar active regions. 
If a vector magnetogram is available, the vertical field $B_z$ on $\cs$
is provided by rotating the measured magnetic field components to the local, 
heliographic, reference system (Gary \& Hagyard 1990). 
Alternatively, the line-of-sight component can be used instead of
$B_z$ provided that the studied active region is 
located sufficiently close to the center of the solar disk. 
This requirement typically minimizes the impact of viewing projection
effects caused by the curvature of the solar surface. 

Unlike the potential magnetic energy $E_p$, the total magnetic 
energy $E$, equation (\ref{ten2}), 
cannot be calculated without additional assumptions or
by using line-of-sight magnetograms. This is because $\mbf{A}$ and 
$\nabla \times \mbf{B}$ are generally unknown on and above the
(photospheric or chromospheric) boundary $\cs$.  
\section{The energy-helicity formula in the linear force-free approximation}
In the force-free approximation, $(\nabla \times \mbf{B}) \times \mbf{B}=0$, 
the total magnetic energy $E$ of a structure extending in $\cv$ 
is provided by the magnetic Virial theorem (Molodensky 1974; Aly 1984) 
\beq
E = {{1} \over {4 \pi}} \int _{\partial \cv} 
[{{1} \over {2}} B^2 \mbf{R} - (\mbf{B} \cdot \mbf{R}) \mbf{B}] \cdot
\mbf{\hat{n}} d \sigma\;\;, 
\label{vir1}
\eeq
where $\mbf{R}$ is a vector position with arbitrary origin in $\cv$.
For planar geometry, $\cv$ extending to infinity and being bounded
only by $\cs$ at $z=0$, and under the 
assumption that the magnetic field strength $B$
vanishes with distance more rapidly than $R^{-3/2}$, 
the Virial theorem reduces to its well-known form 
\beq
E= {{1} \over {4 \pi}} \int _{\cs} \mbf{r} \cdot \mbf{B} B_z d \cs\;\;,
\label{vir}
\eeq
where $\mbf{r}=x \mbf{\hat{x}} + y \mbf{\hat{y}}$ 
is a vector position with arbitrary origin on $\cs$. 
Equation (\ref{vir}) has been applied to solar active regions 
(Metcalf, Leka, \& Mickey 2005; Wheatland \& Metcalf 2006) 
assuming potential or force-free (not necessarily linear) magnetic
fields. 
The explicit dependence of equation (\ref{vir}) on the coordinate system 
leads to inconsistencies if the employed magnetic field vector is not 
force-free (for a detailed discussion of problems related to the
magnetic Virial theorem see Klimchuk, Canfield, \& Rhoads 1992). 
Although well-known and particularly useful, the Virial theorem does not  
link the magnetic energy budgets with the relative
magnetic helicity in a self-consistent way. 
For this reason, we will hereafter 
follow our alternative formulation for the potential magnetic
energy (equation (\ref{pen1})) and the total magnetic energy in the
LFF approximation. As shown in Figure \ref{ecomp} and
explained in \S5.2, the Virial theorem and our energy expressions give
very similar results. 

Implementing the force-free approximation, 
$\nabla \times \mbf{B}= \alpha \mbf{B}$, the total magnetic energy
from equation (\ref{ten2}) gives 
\beq
E=
{{1} \over {8 \pi}} \int _{\cs} \mbf{B} \times \mbf{A_p} \cdot \mbf{\hat{z}} \;d \cs
+ {{1} \over {8 \pi}} \int _{\cv} \alpha \mbf{A} \cdot \mbf{B} \;d \cv\;\;,
\label{ten3}
\eeq
and corresponds to the {\it energy-helicity formula} of Berger (1988). 
In case of the LFF approximation, where the force-free
parameter $\alpha$ is constant in $\cv$, the dependence between $E$
and the relative magnetic helicity $H_m$ becomes explicit. 
Substituting equation (\ref{hm}) into equation (\ref{ten3}) 
for constant $\alpha$, we obtain 
\beq
E= {{1} \over {8 \pi}}
\int _{\cs} \mbf{B} \times \mbf{A_p} \cdot \mbf{\hat{z}} \; d \cs + 
{{\alpha} \over {8 \pi}} H_m\;\;.
\label{ehf1}
\eeq 
Since the relative magnetic helicity depends entirely on the presence
of electric currents so that $H_m=0$ for $\mbf{B}=\mbf{B_p}$, the
first term in the rhs of equation (\ref{ehf1}) must correspond to the
magnetic energy that does not include the energy stored in electric
currents for any {\it nonzero} $\alpha$ and $H_m$. This ground-state
energy can only be the potential energy $E_p$, so the energy-helicity
formula in the LFF approximation reads
\beq
E=E_p + {{\alpha} \over {8 \pi}} H_m\;\;.
\label{ehf}
\eeq 
A proof of equation (\ref{ehf}) is provided in Appendix A. 

In a constant-alpha magnetic structure 
the {\it sense} (sign) of the relative magnetic helicity $H_m$ is 
dictated by the {\it chirality} 
(sign) of the unique value of the force-free parameter $\alpha$. Therefore, 
$\alpha H_m >0$ by definition in the LFF approximation, 
where $\alpha$ and $H_m$ are nonzero.
If $\alpha H_m=0$, then both $\alpha$ and $H_m$ are zero by definition. In this 
case, $E=E_p$ from equation (\ref{ehf}). 
By means of equation (\ref{eneq}), moreover, the LFF approximation
implies a linear dependence 
between the free magnetic energy $E_c$ and the relative magnetic 
helicity $H_m$, namely, 
\beq
E_c={{\alpha} \over {8 \pi}} H_m\;\;.
\label{npen}
\eeq
The monotonic dependence of $E_c$ on $\alpha$ and $H_m$ can be
understood if one considers that both a nonzero relative helicity and
a nonzero free energy depend on, and are directly proportional to, the
existence of electric currents ($\alpha \ne 0$). Of course, equation
(\ref{npen}) is valid only for constant $\alpha$ because the sense of
helicity is the same throughout the magnetic structure. For a
non-constant $\alpha$ and in case of equal and opposite amounts of
helicity being present in the structure, the net relative helicity
$H_m$ becomes zero. This would give $E_c=0$ for $\alpha \ne 0$ in 
equation (\ref{npen}), which is not true. 

Given that the potential energy is readily calculated (equation
(\ref{pen1})), it is evident from equation (\ref{ehf}) that 
knowledge of the relative magnetic helicity $H_m$ is sufficient to 
fully evaluate the LFF energy-helicity formula. 
However, $H_m$ cannot be evaluated 
from the general equation (\ref{hm}). This is because the vector 
potential $\mbf{A}$ is unknown in $\cv$, although the LFF magnetic field $\mbf{B}$ 
can, in principle, be calculated everywhere in $\cv$. In the next section we 
derive convenient expressions for $\mbf{A}$ and $H_m$ in the LFF approximation. 
\section{The relative magnetic helicity in the linear force-free approximation}
\subsection{Volume-integral representation}
It is straightforward to obtain a volume-integral expression for the
relative magnetic helicity $H_m$ from the energy-helicity formula,
equation (\ref{ehf}), using the definitions of the potential and the
total magnetic energies: 
\beq
H_m = {{1} \over {\alpha}} \int _{\cv} (B^2 - B_p^2) d \cv\;\;.
\label{hm_hs}
\eeq
This expression was used by Hagino \& Sakurai (2004) who assumed
$\mbf{A}=0$ on $\cs$. Despite its simplicity, however, equation
(\ref{hm_hs}) cannot be directly compared to the general equation 
(\ref{hm}) because the form of $\mbf{A}$ is not obvious. To make this
conceptual step, we decompose $\mbf{B}$ in equation (\ref{hm_hs}) into
its potential and nonpotential components to find, after some algebra,
that 
\beq
H_m={{1} \over {\alpha}} \int _{\cv} \mbf{B_c} \cdot \mbf{B} d\cv\;\;,  
\label{hmv}
\eeq
where the condition $\mbf{B_p} \cdot \mbf{B_c}=0$ has been
used. Equation (\ref{hmv}) is identical to equation (\ref{hm})
for $\mbf{A} = (1/\alpha) \mbf{B_c}$, or 
\beq
\mbf{A} = {{1} \over {\alpha}} (\mbf{B} - \mbf{B_p})\;\;.
\label{vp}
\eeq

One may verify that the definition of $\mbf{A}$ in the LFF
approximation, equation (\ref{vp}), complies with all the conditions
of equations (5)-(7). Moreover, it is clear that 
$\mbf{A}|_{\cs} \ne 0$, contrary to what Hagino \& Sakurai (2004)
assumed. This being said, $\mbf{A}$ does not have a vertical component
on $\cs$ because $\mbf{B}$ and $\mbf{B_p}$ share
the same vertical component on $\cs$, equation 
(\ref{bcon2}). Adopting $\mbf{A} \cdot \mbf{\hat{z}} |_{\cs}=0$,
however, equations (9) will give
\beq
\mbf{A_c} \times \mbf{\hat{z}}|_{\cs} = 
\mbf{A_c} \cdot \mbf{\hat{z}}|_{\cs} = 0\;\;.
\label{ac1}
\eeq
From equations (\ref{ac1}), then, $\mbf{A_c}|_{\cs}=0$, rather than 
$\mbf{A}|_{\cs}=0$. Since 
$\mbf{A_c}|_{\cs}=0$, $\mbf{A_p}$ and $\mbf{A}$ coincide on
$\cs$ via equation (\ref{bcon4}), namely
\beq
\mbf{A} |_{\cs} = \mbf{A_p} |_{\cs}\;\;.
\label{aap}
\eeq
Of course, equation (\ref{aap}) does not preclude 
$\nabla \times \mbf{A}|_{\cs} \ne \nabla \times \mbf{A_p}|_{\cs}$. 
This is because these curls include vertical gradient terms 
$(\partial / \partial z)$ and 
$\mbf{A} \ne \mbf{A_p}$ above $\cs$.

Equation (\ref{hmv}) for the relative magnetic helicity satisfies all the 
requirements of the LFF approximation. Moreover, it is 
more complete than the equation 
$H_m=(1/\alpha) \int _{\cv}B^2 d \cv$ of 
Pevtsov, Canfield, \& Metcalf (1995). Clearly, from the above expression 
$\lim _{\alpha \rightarrow 0} |H_m|=E_p \lim _{\alpha \rightarrow 0} (1/|\alpha|)$, 
which tends to infinity while it should tend to zero. Given that $\mbf{B}$ in 
equation (\ref{hmv}) corresponds to an LFF magnetic field, the
integrand 
$\mbf{B_c} \cdot \mbf{B} = B^2 - \mbf{B} \cdot \mbf{B_p}$ 
is known at any location 
in $\cv$ if $B_z$ is known and flux-balanced on $\cs$. 
Therefore, the total relative magnetic helicity of a
closed and flux-balanced magnetic structure with constant $\alpha$ in $\cv$
can be calculated using equation (\ref{hmv}). 
By extension, equation
(\ref{hmv}) can be used to calculate the total relative magnetic
helicity of an isolated solar active region for which the constant-alpha
approximation is assumed valid and for which flux-balanced
photospheric or chromospheric vector magnetic field measurements
exist. A representative, unique value of the force-free parameter $\alpha$ 
can be calculated by an array of techniques (Leka \& Skumanich 1999;
Leka 1999) with an alternative technique described in \S5.1.

In theory, one may estimate the relative magnetic helicity from 
the volume integral of equation (\ref{hmv}). In practice, however, 
it would be both 
risky and computationally extensive to apply equation (\ref{hmv}) to 
extrapolations of actual solar magnetic configurations. This is
because (i) despite the treatment of periodic effects on the
horizontal plane, Fourier transforms in the LFF approximation can
still cause periodic effects in the vertical direction and at large
heights above the photosphere (i.e., Alissandrakis 1981; Gary 1989), 
and (ii) one does not know {\it a priori} the height above the photosphere 
where integration should stop. There are some partial remedies for both 
of the above problems: the use of a much larger 
computation volume than that required to 
contain the magnetic structure may limit periodic effects within $\cv$, 
while the maximum integration height can be either equal to the linear size of the 
surface $\cs$ or determined by the contribution to the relative magnetic helicity 
$H_m$. If integration above a certain height makes insignificant contributions to 
$H_m$ then integration stops at this height. In any case, calculating $H_m$ from 
equation (\ref{hmv}) is a very time-consuming task. For this reason,
we derive in the next section a first-principles {\it surface} integral
for $H_m$ in the LFF approximation. 
\subsection{Surface-integral representation based on Fourier transforms}
Regardless of the force-free approximation, the total magnetic energy
$E$ of a closed magnetic structure extending into $\cv$ and rooted in $\cs$
consists of the potential magnetic energy $E_p$ of the structure
and the nonpotential (free) magnetic energy $E_c$ due to electric
currents (equation (\ref{eneq})). Expressing $E_c$ in terms of $E_p$,
one may write 
\beq
E=(1+ f)E_p\;\;,
\label{es1}
\eeq
where $E_c=f E_p$ and $f$ is generally a positive and dimensionless variable.
The constant-alpha approximation readily provides a
condition for $f$, namely $\lim _{\alpha \rightarrow 0} f =0$. In
addition, $f$ must be a function of $\alpha$ and increasing $|\alpha|$
should increase $f$ monotonically giving rise to 
a symmetric profile of $f(\alpha )$ with respect to 
$\alpha =0$, i.e. $f(|\alpha|)=f(-|\alpha|)$. The form of $f$ can be
derived analytically in the LFF approximation if one uses the
formulation of Berger (1985). The details of the derivation are given
in Appendix B. Here we provide two expressions for the variable
$f$. The first is the exact analytical formula, while the second is a
linearized, with respect to $\alpha ^2$, version $f_l$ of it, 
useful to keep the free energy and relative helicity finite when 
$|\alpha| d \rightarrow (2 \pi /L)$, where $d$ is the elementary size
on the boundary $\cs$ and $L$ is the linear size of the magnetic
structure on $\cs$. For an observed magnetogram, $d$
corresponds to the linear size of a pixel expressed in physical units.  
In particular, 
\beq
f = \mathcal{F} \;\;,\;and\;\;\;\;\; f_l = \mathcal{F}_l d^2 \alpha ^2\;\;,
\label{fn}
\eeq
where 
\begin{mathletters}
\beq
\mathcal{F}={{\sum _{l=1}^{n_x} \sum _{m=1}^{n_y} |b_{u_l,v_m}^2| 
             {{(u_l^2 + v_m^2)^{1/2} - (u_l^2 + v_m^2 - \alpha ^2 d^2)^{1/2}}
             \over {(u_l^2 + v_m^2)^{1/2} (u_l^2 + v_m^2 - \alpha ^2d^2)^{1/2}}}}
             \over 
             {\sum _{l=1}^{n_x} \sum _{m=1}^{n_y} 
             {{|b_{u_l,v_m}^2|} \over {(u_l^2 + v_m^2)^{1/2}}}} }\;\;,
\eeq
and
\beq
\mathcal{F}_l={{1} \over {2}}{{\sum _{l=1}^{n_x} \sum _{m=1}^{n_y} 
                {{|b_{u_l,v_m}^2|} \over {(u_l^2 + v_m^2)^{3/2}}}} 
               \over 
               {\sum _{l=1}^{n_x} \sum _{m=1}^{n_y} 
                {{|b_{u_l,v_m}^2|} \over {(u_l^2 + v_m^2)^{1/2}}}}}\;\;,
\eeq
\end{mathletters}
respectively. In equations (34), $b_{u_l,v_m}$ is the Fourier
amplitude of the vertical magnetic field $B_z$ for the harmonic
$(u_l,v_m)$ in a Fourier space with dimensions $n_x$, $n_y$.  
The linearization $f_l$ implies a minimum value for $f$ that results in the
estimation of a minimum free magnetic energy $E_c$ and relative
magnetic helicity $|H_m|$ in the LFF approximation. 
The underestimation of $E_c$ and $H_m$ is
negligible for small $|\alpha|$ and increases as $\alpha$ increases 
(see \S4.3 below and Figure \ref{hc}). This effect is explained in
detail in Appendix B and has been realized by several previous
works when the linearized relative helicity expression of
equation (\ref{in2}) was derived 
(Green et al. 2002; D\'{e}moulin et al. 2002;
D\'{e}moulin 2006). The infinite energy and helicity for 
$(|\alpha|/d) \rightarrow (2 \pi/L)$ 
is a well-known problem of the LFF magnetic fields that are not fully
described by the boundary condition on $\cs$ in this case 
(e.g., Alissandrakis 1981).    

The quadratic dependence of $f$ on $\alpha$ in both the exact and the
linearized case guarantees a 
symmetric profile of $f(\alpha)$ and a vanishing $f$ for 
$\alpha \rightarrow 0$. This dependence has also been demonstrated
graphically by Sakurai (1981) in analytical force-free fields. 

From equations (\ref{es1}) and (\ref{fn}) we can now parameterize all terms 
of the energy-helicity formula (equation (\ref{ehf})) with respect to one of 
these terms. As the free parameter we choose the potential magnetic energy $E_p$ 
for which the general expression of equation (\ref{pen1}) exists. 
Then, the exact and the linearized surface-integral expressions for
the total magnetic energy in a constant-alpha magnetic structure read
\begin{mathletters}
\beq
E=(1 + \mathcal{F}) E_p\;\;,
\eeq
and 
\beq
E=(1 + \mathcal{F}_l d^2 \alpha ^2) E_p\;\;, 
\eeq
\end{mathletters}
respectively. For the free magnetic energy we obtain 
\begin{mathletters}
\beq
E_c=\mathcal{F} E_p\;\;,
\eeq
and
\beq
E_c=\mathcal{F}_l d^2 \alpha ^2 E_p\;\;,
\eeq
\end{mathletters}
respectively, while for the relative magnetic helicity we find 
\begin{mathletters}
\beq
H_m = {{8 \pi} \over {\alpha}} \mathcal{F} E_p\;\;,
\eeq
and
\beq
H_m = 8 \pi \mathcal{F}_l d^2 \alpha E_p\;\;,
\eeq
\end{mathletters}
respectively. Notice that the
exact formula for the relative helicity, equation (37a), still yields
$|H_m| \rightarrow 0$ for $|\alpha| \rightarrow 0$ because
$\mathcal{F} \propto \alpha ^2$ tends to zero faster than $\alpha$. 
Equation (37b) 
gives values that are a factor of four smaller than those of the
linearized relative helicity of D\'{e}moulin et al. (2002) and Green et
al. (2002) (equation (\ref{in2})). This can be seen 
from equations (37b) and (\ref{apc3}), and by setting 
$d=1/(2 \pi)$, which corresponds to Berger's (1985) unit length
assuming a computational box with linear size $L$ equal to unity. Part of
the discrepancy has been corrected by D\'{e}moulin (2006) who admits that
the original expression of equation (\ref{in2}) was a factor of two
too high. In Appendix B we show that the linearization 
introduces an additional $(1/2)$-factor in equation (34b). 

Concluding, equations (35) - (37) offer convenient 
surface-integral representations of the relative magnetic helicity, 
as well as of the total and the nonpotential magnetic energies in the LFF 
approximation. Equations (\ref{pen1}) and (35)-(37) 
will be evaluated in \S\S4.3 and 5 for semi-analytical and observed
magnetic configurations, respectively. 

From equations (35) - (37) we can calculate the uncertainties 
$\delta E$, $\delta E_c$, and $\delta H_m$ of the total magnetic 
energy, the free magnetic energy, and the total magnetic helicity,
respectively. Uncertainties of the potential
energy, equation (\ref{pen1}), stem from the uncertainties 
$\delta B_z$ of the
normal (vertical) magnetic field component $B_z$. Although 
the values of $\delta B_z$ are generally known for
a given magnetogram, it is difficult
to propagate them into the potential energy because of the
extrapolations required to infer the potential magnetic field and its
vector potential. In our case the extrapolations are performed using
Fast Fourier transforms. For this reason we will ignore the uncertainties
$\delta E_p$ of the potential energy, although we expect that these
uncertainties should not be very significant, given that the
vertical magnetic field component is the least uncertain measured
component of the magnetic field vector, especially for active regions
located close to the center of the solar disk. 
For the same reason we will also ignore the uncertainties 
$\delta \mathcal{F}_l$ of $\mathcal{F}_l$. Excluding $\delta E_p$ and 
$\delta \mathcal{F}_l$, the only source of uncertainties is the
uncertainties $\delta \alpha$ in the inference of the force-free
parameter $\alpha$. These uncertainties give rise to a nonzero  
$\delta \mathcal{F}$ in the value of $\mathcal{F}$ (equation
(34a)). From equations (35) - (37), then, we obtain the following
uncertainty expressions: 

For the exact and linearized total magnetic energy, 
\begin{mathletters}
\beq
{{\delta E} \over {E}} = {{E_c} \over {E}} {{\delta \mathcal{F}} \over
  {\mathcal{F}}}\;\;,
\eeq
and
\beq
{{\delta E} \over {E}} = 2 {{E_c} \over {E}} {{\delta \alpha} \over
  {|\alpha|}}\;\;,
\eeq
\end{mathletters}
respectively. For the free magnetic energy, 
\begin{mathletters}
\beq
{{\delta E_c} \over {E_c}} = {{\delta \mathcal{F}} \over {\mathcal{F}}}\;\;,
\eeq
and 
\beq
{{\delta E_c} \over {E_c}} = 2 {{\delta \alpha} \over {|\alpha|}}\;\;,
\eeq
\end{mathletters}
respectively. For the relative magnetic helicity, 
\begin{mathletters}
\beq
{{\delta H_m} \over {|H_m|}} \le \sqrt{({{\delta \alpha} \over
                                       {\alpha}})^2 + 
                             ({{\delta \mathcal{F}} \over {\mathcal{F}}})^2}\;\;,
\eeq
and 
\beq
{{\delta H_m} \over {|H_m|}} = {{\delta \alpha} \over {|\alpha|}}\;\;,
\eeq
\end{mathletters}
respectively. The ``$\le$'' symbol in equation (40a) is due
to the fact that $\mathcal{F}$ and $\alpha$ are interrelated. 
Given that it is also difficult to propagate the uncertainties
of $|\alpha|$ into $\delta \mathcal{F}$, we will hereafter use the
linearized expressions of the uncertainties, equations (38b) - (40b). 
\subsection{Comparison between the volume- and the surface-integral expressions
for the relative magnetic helicity}
At this point we have derived two types of 
expressions for the relative magnetic helicity in the 
LFF approximation, namely the volume integral of equation (\ref{hmv}) and the 
surface integrals of equations (37). To ensure consistency, 
these expressions must provide nearly 
identical results for small values $|\alpha|$ of the force-free
parameter, while the linearized expression of equation (37b) must
provide a lower limit of the relative magnetic helicity as $|\alpha|$
increases. To avoid errors due to
observational uncertainties and to make a safer evaluation of the volume
integral of equation (\ref{hmv}) we use 
semi-analytical models of magnetic
structures, rather than observed solar magnetograms. 
For a simple representation of 
the twist present in the magnetic configurations 
we use dipolar magnetic field models. 
For a given dipole with footpoint separation $L_{sep}$, 
we define the dimensionless quantity $N = \alpha L_{sep}$. This
quantity is generally a dimensionless measure of $\alpha$. In the
particular case of field lines winding about an axis (not necessarily
assumed here), $N$ is a measure of the total end-to-end number of
turns of the dipole. Our dipoles are characterized according to their
$N$-values. We first create the analytical distribution for the
vertical magnetic field $B_z$ normal to the horizontal plane $\cs$ and
then we apply LFF extrapolations in the volume $\cv$ using
the same $B_z$-distribution as boundary condition and assuming
different $\alpha$-values stemming from different $N$-values in each
extrapolation. Extrapolations are
performed using the Fast Fourier transform method of Alissandrakis
(1981). For this test we use positive $\alpha$-values which results in
right-handed helicities. Using equal and
opposite $\alpha$-values would only change the sense of twist and
hence the sign, but not the magnitude, of the calculated magnetic
helicity. The magnetic energy budgets, equations (35) - (36), are
insensitive to the sign of $\alpha$. 

Our model dipoles have a fixed footpoint separation length
$L_{sep}=100$ (in arbitrary units) and are embedded in a boundary plane
$\cs$ with linear dimensions $L_x = L_y = 200$, assuming an elementary
length $d=1$ as the unit length. The separation length $L_{sep}$ 
represents the distance between the positive- and the
negative-polarity centers of the dipole on
$\cs$. We use an array of $N$-values, where $N \in [0.02, 2]$, 
each of which determines a different $\alpha=(N/L_{sep})$. 
To make sure that equations (35) - (37) do not depend on the details 
of a particular model, we use three different models of $B_z$ on
$\cs$. For each model, the two polarity centers 
are placed at vector positions $\mbf{r_1}$ and $\mbf{r_2}$,
respectively, on $\cs$, such that $|\mbf{r_1} - \mbf{r_2}| = L_{sep}$. 
The number of harmonics used for the Fourier-transform 
calculation of $\mathcal{F}$ and $\mathcal{F}_l$, equations (37a) and
(37b), respectively, is 
kept fixed in all cases and is equal to $n_x=n_y=256$. 
We use the following models: 
\ben
\item[(1)] A Gold-Hoyle dipole solution (Gold \& Hoyle 1960), i.e., 
\beq
B_z(\mbf{r})=B_0 [{{1} \over {1+q(\mbf{r}-\mbf{r_1})^2}} -  {{1} \over
    {1+q(\mbf{r}-\mbf{r_2})^2}}]\;\;,
\label{gh}
\eeq
where $\mbf{r}$ is the vector position of a given location on $\cs$ 
and $B_0$, $q$ are positive
constants. In this test we have used a fixed $B_0=10^3$ and an array
of $q$-values, $q \in [0.5, 10]$. Each $q$-value has been applied to
the full array of $N$-values. 
\item[(2)] A solenoidal dipole solution (Sakurai \& Uchida 1977), i.e., 
\beq
B_z (\mbf{r})= {{B_0} \over {16 \pi}} 
(16 + \pi ^2 \bar{L}_{sep}^2)^{3/2} \sum _{i=1} ^2 \{
{{s_i} \over {\sqrt{16(1+\bar{\rho}_i)^2 + \pi ^2 \bar{L}_{sep}^2}}} 
[I_1(k_i) + {{16(1-\bar{\rho}_i^2) - \pi ^2 \bar{L}_{sep}^2} \over 
{16(1-\bar{\rho}_i)^2 + \pi ^2 \bar{L}_{sep}^2}} I_2(k_i)] \}\;\;,
\label{su}
\eeq
where $s_1=1$, $s_2=-1$, $q$ is a positive constant, 
$\bar{\rho _i}=(\rho _i /q)$; $i \equiv \{ 1,2 \}$, are the
normalized, with respect to $q$, distances of a location $\mbf{r}$ on
$\cs$ from $\mbf{r_1}$ and $\mbf{r_2}$ 
($\rho _i = |\mbf{r} - \mbf{r_i}|$; $i \equiv \{ 1,2 \}$),
$\bar{L}_{sep}=L_{sep}/q$ is the normalized, with respect to $q$,
separation length, 
$k_i^2= 64 \bar{\rho}_i / [16(1+\bar{\rho}_i)^2 + \pi ^2 \bar{L}_{sep}^2]$, 
and $I_1(k_i)$, $I_2(k_i)$ are the complete 
elliptic integrals of the first and second kind, respectively, i.e., 
\beq
I_1(k_i)= \int _0 ^{\pi/2} {{d \theta} \over {\sqrt{1 - k_i^2 sin^2 \theta}}}
\;\;\;\;and\;\;\;\;
I_2(k_i)=\int _0 ^{\pi/2} \sqrt{1 - k_i^2 sin^2 \theta} d \theta\;\;.
\label{elin}
\eeq
Here we have used a fixed $B_0=10^3$ and an array
of $q$-values, $q \in [0.02, 0.3]$, each value of which has been
applied to the full array of $N$-values.  
\item[(3)] A submerged poles dipole solution (Longcope 2005 and
references therein), i.e., 
\beq
B_z (\mbf{r})= B_0 q \{ 
{{1} \over {[(\mbf{r} - \mbf{r_1})^2 +q^2]^{3/2}}} - 
{{1} \over {[(\mbf{r} - \mbf{r_2})^2 +q^2]^{3/2}}} \}\;\;,
\label{sp}
\eeq     
where $B_0$ and $q$ are positive constants. The constant $q$, in
particular, represents the depth below $\cs$ in which the two magnetic
monopoles are placed. The depth of each monopole can, in principle,
be different than that of the other(s), but here we use a fixed
depth, as well as a fixed magnetic field strength $B_0$ for each
monopole, to create a flux-balanced magnetic configuration on $\cs$. 
Here we use $B_0=10^3$ and an array of depths $q$, $q \in [0.5, 10]$, 
each value of which has been applied to the full array of $N$-values.  
\een

Comparing the volume-integral expression, equation (\ref{hmv}), with
the surface-integral expressions, equations (37), for the relative
magnetic helicity $H_m$ gives the expected results for all $q$- and
$N$-values. Three of these results, one for each model, are given in
Figure \ref{hc}a. The $q$-values for each model in Figure \ref{hc}a 
were selected with the
sole purpose of giving rise to well-separated helicity values, for
convenience in the visual comparison. These selections are
$q=0.6,\;0.2$, and $10$ for the Gold-Hoyle (GH), the Sakurai-Uchida
(SU) and the submerged-poles (SP) model, respectively. Figure
\ref{hc}a gives rise to the following conclusions:
\ben
\item[(1)] All helicity expressions give very similar results for a
  given model, which suggests that the LFF equations (35) - (37) are
  model-independent. 
\item[(2)] For $N= \alpha L_{sep} \rightarrow 0$, all expressions give
  $H_m \rightarrow 0$. Therefore, $H_m$ from equations (\ref{hmv}) and
  (37) corresponds to the gauge-invariant relative magnetic helicity
  discussed in \S\S 4.1 and 4.2. As $\alpha$ increases for a fixed
  $L_{sep}$, Figure \ref{hc}a shows the quadratic increase of the
  magnetic helicity in the dipoles, for a fixed boundary condition
  $B_z$ on $\cs$. 
\item[(3)] Clearly, all expressions for $H_m$ give nearly identical
  results for small $\alpha$. As $\alpha$ increases, the linearized
  surface-integral expression (equation (37b); dotted curves)
  consistently provides a lower $H_m$ (the exact surface-integral
  expression of equation (40a) is represented by solid curves and
  rectangles), as expected. The volume-integral expression for 
  $H_m$ (equation (\ref{hmv}); dashed curves and triangles) 
  gives slightly higher values than both surface-integral expressions. 
\een

Tests with $N >2$ (not shown in Figure \ref{hc}) reveal that the
volume-integral $H_m$ increases exponentially after some maximum
$N$-value, while the surface-integral expressions continue to increase
quadratically{\footnote{Of course, at the vicinity of 
$|\alpha| d \simeq (2 \pi /L)$, which in our parameter selection
corresponds to $N = \pi$, the exact surface-integral $H_m$
increases abruptly to become infinite for $|\alpha| d = (2 \pi /L)$}}.
This maximum $N$-value is model-dependent and, in case of the SP
model, it changes even with varying model parameters. This inability
to predict the behavior of the volume-integral $H_m$ for large $N$,
combined with spurious results of the Fourier-transform extrapolations
in these cases, enhances one's impression that the volume-integral
helicity is less reliable than the surface-integral expressions and
that, among other problems, it is susceptible to artifacts incurred by
Fourier-transform extrapolations at large heights above the boundary
$\cs$. 

Figure \ref{hc}b shows the underestimation factor caused by the use of
the linearized surface-integral expression of equation (37b) for the
three cases depicted in Figure \ref{hc}a. In
particular, the dashed curves show the ratio between the
volume-integral and the linearized surface-integral helicities, while
the solid curves show the ratio between the exact surface-integral and
the linearized surface-integral helicities. Evidently, the
underestimation 
factor is nearly model-independent when the two surface-integral
expressions are compared, while the situation is less predictable
when the volume-integral and the linearized surface-integral
expressions are compared. From the comparison between the
surface-integral helicity expressions, one sees
that, even for large $N$-values, underestimation does not exceed a
factor of $\sim 1.85$. For dipolar solar active regions for which the
LFF approximation is assumed and with typical values of 
$\alpha \sim 10^{-2}\;Mm^{-1}$ and $L_{sep} \sim 100\;Mm$, we obtain
$N \sim 1$. The expected underestimation factor for this case is
$\lesssim 1.1$, which is very modest compared to the errors expected
from other assumptions, and especially the use of the
constant-alpha approximation itself. 

In summary, the results shown in Figure \ref{hc} demonstrate that
the surface-integral expressions of equations (37) lead to
reliable estimates of the total relative magnetic helicity in a 
constant-alpha magnetic structure. 
By extension, the surface integrals of equations
(35) and (36) provide reliable estimates of the total
magnetic energy and the free magnetic energy, respectively, in the
structure. 
\section{Application to observed solar active region magnetic fields}
\subsection{Data selection and determination of basic parameters}
In this section we apply the results of the previous analysis to
vector magnetograms of solar active regions. In particular,
we calculate the LFF magnetic energy and helicity budgets 
(equations (\ref{pen1}) and (35) - (37)) using  
photospheric vector magnetogram data obtained by the Imaging Vector
Magnetogram (IVM; Mickey et al. 1996; 
LaBonte, Mickey, \& Leka 1999) of the University of Hawaii's
Mees Solar Observatory. IVM's photospheric 
magnetography{\footnote{Recently, the IVM focused on 
the chromospheric magnetically sensitive line 
Na {\small I} ($5896$ \AA). These observations have started providing 
{\it chromospheric} vector magnetograms of solar active regions.}} 
consists of recording the complete Stokes vector at
each of $30$ spectral points through the Fe {\small I} $6302.5$ 
\AA$\;$photospheric spectral line. The line-of-sight magnetic field
components are obtained
via the inversion code of Landolfi \& degl'Innocenti (1982) that
includes LTE radiative transfer, magneto-optic effects, and the
filling factor of the unresolved flux tubes. Our equations are
applicable to the heliographic magnetic field components on the
heliographic plane. The required coordinate transformation
has been carried out following the analysis of Gary \& Hagyard
(1990). 

To test our derivations for energy and helicity, we study two solar
active regions (ARs): a small, short-lived emerging flux region (NOAA
AR 8844) and a persistent, large, and complex AR that exhibited
significant eruptive activity (NOAA AR 9165). For both ARs, the IVM
recorded a series of vector magnetograms over a period of a few 
hours that can be used to follow the temporal evolution of the
energy and helicity budgets in the regions. 

Before applying any analysis that employs the transverse field of 
a vector magnetogram, one must first
resolve the intrinsic azimuthal ambiguity of $180^o$ in the
orientation of the transverse magnetic field component. Azimuth
disambiguation of the employed IVM magnetograms was performed by means
of the nonpotential magnetic field calculation (NPFC) method of
Georgoulis (2005a) - see also Metcalf et al. (2006) for a comparative
evaluation of the method with respect to other disambiguation
methods. Figures \ref{8844}a and \ref{8844}b depict two disambiguated
vector magnetograms of NOAA ARs 8844 and 9165, respectively. Only part
of the IVM field of view is shown in both images, to exemplify
the magnetic structure of the two ARs. Shown are the heliographic
magnetic field components on the heliographic plane. The relative
isolation of the two ARs on the solar disk at the time of the IVM
observations (2000 January 25 and September 15 for ARs 8844 and 9165,
respectively), as well as the ARs' very different records of eruptive
activity prompted us to use these data in this first test of our LFF
energy / helicity calculations. 

The timeseries of the magnetic flux $\Phi$ during the IVM observing
interval for both ARs are shown in Figure \ref{flen_fig}. For both
cases, we notice that the IVM field of view encloses fairly
well-balanced magnetic flux distributions. NOAA AR 8844 is more
flux-balanced than NOAA AR 9165, with an imbalance always kept below
$5$\%. The maximum imbalance of NOAA AR 9165 is around
$10$\%. Our derivations require flux-balanced magnetic structures and
the above slight imbalances are
not expected to significantly impact our results. The first noticeable 
difference between the two ARs is in their respective amounts of
magnetic flux: on average, the magnetic flux in NOAA AR 9165 
($\Phi \sim 17.1 \times 10^{21}\;Mx$) is a factor of $\sim 3.4$ larger
than the flux in NOAA AR 8844 ($\Phi \sim 5.1 \times 10^{21}\;Mx$).
One might also notice a very slight
increasing trend in the evolution of $\Phi$ in NOAA AR 8844 
(from $\sim 4.9 \times 10^{21}\;Mx$ to $\sim 5.3 \times 10^{21}\;Mx$)
within the $2\;hr$ of the IVM observations, implying that the magnetic
structure is growing. This is typical of emerging flux regions. 

After disambiguation, 
we need to calculate a unique force-free parameter $\alpha$ 
for each magnetogram. To do so, we calculate the slope in the
scatter plot between the vertical curl $(\nabla \times \mbf{B})_z$ of the
magnetic field $\mbf{B}$ and the vertical field $B_z$ for strong-field
locations of the magnetograms. By strong-field locations we mean 
locations with magnetic field components exceeding the $1 \sigma$
threshold, where we have taken $1 \sigma$ to 
correspond to a vertical magnetic field of $100\;G$ 
and a horizontal magnetic field of $200\;G$, typical of the IVM. 
Because the LFF approximation is a
gross simplification of the photospheric active-region magnetic
fields, however, the uncertainty in the value of the slope is often  
larger than the slope itself due to the substantial scatter in the
pairs of $[(\nabla \times \mbf{B})_z\;,\;B_z]$-values. To restrict the 
uncertainty in the calculation of $\alpha$ we have developed 
the following procedure: we obtain several $\alpha$-values from the
slope of the scatter plot, each calculated using a different
significance threshold, as shown in Figure \ref{acalc}. Let $\alpha _k$ be the
value of $\alpha$ for a given significance threshold 
$\sigma _k = k \sigma$; $k \ge 1$ 
and $\Phi _k$ be the respective unsigned magnetic flux used 
in the calculation. Then, the adopted unique value of $\alpha$
and its uncertainty $\delta \alpha$ are obtained by the flux-weighted
averages 
\beq
\alpha = {{\sum _k \alpha _k \Phi _k} \over {\sum _k \Phi _k}}
\;\;\;\;and\;\;\;\;
\delta \alpha = {{\sum _k |\alpha - \alpha _k| \Phi _k} \over {\sum _k
                  \Phi _k}}\;\;,
\label{fla}
\eeq
respectively. An example of this calculation is shown in Figure
\ref{acalc}. 
The flux-weighted average $\alpha$ is indicated 
by the solid line and the surrounding shaded area
indicates the extent of its uncertainty $\delta \alpha$. The above
process provides a maximum-likelihood $\alpha$-value with a reasonable
uncertainly and is repeated for every vector magnetogram of the
timeseries to obtain the respective timeseries for $\alpha$. 
Using other methods to calculate $\alpha$ 
(see, e.g., Leka \& Skumanich 1999 and Leka 1999) we verified that the
timeseries of $\alpha$ obtained by equations (\ref{fla}) are more smooth
(less spiky) and with smaller uncertainties for each $\alpha$-value,
than the $\alpha$-timeseries stemming from the other methods. 
\subsection{Magnetic energy and helicity calculations}
The timeseries of the force-free parameter $\alpha$ for both tested
ARs are shown in Figure \ref{alp}. As we discussed in \S5.1, the
different $\alpha$-values are generally consistent with each other,
giving rise to fairly well-defined averages in both cases. The overall
twist for NOAA AR 8844 is right-handed ($\alpha >0$), while for NOAA
AR 9165 it is left-handed ($\alpha <0$). For the latter AR, in
particular, $\alpha$ appears to decrease, in absolute value, in the
course of time. Coincidentally, the average absolute values
$|\bar{\alpha}|$ of $\alpha$ for both ARs are almost identical 
($\bar{\alpha} = 0.023 \pm 0.06\;Mm^{-1}$ and 
$\bar{\alpha} = -0.024 \pm 0.06\;Mm^{-1}$ for NOAA ARs 8844 and 9165,
respectively). We note in passing that the value of $\bar{\alpha}$ for NOAA
AR 8844 is in excellent agreement with the value of $0.022\;Mm^{-1}$,
calculated by Pariat et al. (2004). The latter $\alpha$-value was
inferred by combining the best LFF match of the active-region corona
using simultaneous EUV images from TRACE with a best LFF fit of the
observed horizontal magnetic field. The magnetic field vector in
Pariat et al. (2004) was acquired by the high-resolution 
vector magnetograph onboard the balloon-borne Flare Genesis Experiment
(FGE; Bernasconi et al. 2001).  

Although the two studied ARs happen to have almost the same
$\alpha$-values, albeit with different signs, the much larger magnetic
flux carried by the eruptive NOAA AR 9165 is expected to lead to much
larger energy / helicity budgets than the respective budgets of the
noneruptive NOAA AR 8844. The relative magnetic helicity and the
respective magnetic energies for ARs 8844 and 9165 are plotted in
Figures \ref{hen1} and \ref{hen2}, respectively. There we show both
the linearized (equations (35b)-(37b); red curves) and the exact 
(equations (35a)-(37a); blue curves) surface-integral expressions
for energy and helicity. Before discussing and comparing individual
values, we note that the linearized expressions generally provide 
slightly lower magnitudes of energy and helicity. 
This is clearly the case for NOAA AR 8844 (Figure \ref{hen1}), 
while in cases where the linearized values are larger
than the exact values for NOAA AR 9165 (Figure \ref{hen2}), 
the difference is within error
bars. That the linearized expressions provide lower limits of energy
and helicity was concluded from the analysis in Appendix B and
verified using semi-analytical models in \S4.3. 
In both cases of observed ARs, moreover, 
the timeseries of the exact values
appear more spiky than the respective timeseries of the linearized
values. Given that the linearized expressions 
lead to a smoother temporal evolution with slightly
lower values than the exact expressions, 
equations (35b) - (37b) for the linearized 
energy and helicity budgets appear preferable compared to the exact
expressions of equations (35a) - (37a). Besides being more convenient
and well-behaved, the linearized expressions also have readily
derivable uncertainties (equations (38b) - (40b)) based on the
uncertainties $\delta \alpha$ of $\alpha$. This being said, one
notices the close correspondence of the $\alpha$-value timeseries of
Figures \ref{alp} with the timeseries of the linearized 
helicity of Figures \ref{hen1} and \ref{hen2}. Clearly, the success of
the LFF energy/helicity estimations depends on the 
reliability of the inference 
of $\alpha$. This is a key feature that one should
keep when trying to generalize the LFF energy/helicity formulas into
NLFF ones, valid for a variable $\alpha$ within the field of view. 

The average energy/helicity values from Figures \ref{hen1} and
\ref{hen2} are summarized and compared in Tables \ref{Tb1} and
\ref{Tb2}. Table \ref{Tb1} shows the comparison between average
magnetic fluxes, $\alpha$-values, and helicities, while Table
\ref{Tb2} focuses on the comparison between the various average energy
budgets from the two ARs. It is quite useful that the average
$\alpha$-values are nearly identical for the two ARs. We then notice
that the eruptive NOAA AR 9165, with a factor of $\sim 3.4$ larger
magnetic flux than the noneruptive NOAA AR 8844, also has a potential
and a total magnetic energy that are similarly (by a factor of 
$\sim 3.2-3.5$) larger than those of NOAA AR 8844. The average
relative magnetic helicity and free magnetic energy of the eruptive
AR, however, are $\sim 7.6-8.9$ times larger than those of the
noneruptive AR. Notably, the linearized expressions, that result 
in lower values and uncertainties, 
consistently give a higher factor of difference in both energy and
helicity. The much wider difference between the free energy and the
relative helicity between the two ARs suggests that a viable criterion 
(safer than simply evaluating the magnetic flux) for distinguishing
between eruptive and noneruptive ARs 
may be the amount of free magnetic energy and helicity. 
This study should, of course, be applied to a
large number of eruptive and noneruptive ARs for the results to
obtain statistical significance. Moreover, if this treatment is
generalized to the NLFF, rather than the LFF, approximation, it will
be much more physically meaningful given the expected conditions in the
low solar atmosphere. Forced photospheric fields
(see, e.g., Georgoulis \& LaBonte (2004)) should still lead to 
discrepancies stemming from the application of the force-free
approximation. Another notable fact from Table \ref{Tb2}
is the fractional free magnetic energy $(\bar{E}_c/\bar{E})$,
normalized by the total magnetic energy. For the noneruptive NOAA
AR 8844, the free energy is $\sim 4.7$\% - $6$\% of the total magnetic
energy. For the eruptive NOAA AR 9165, the free energy corresponds to 
 $\sim 12.3$\% - $13.1$\% of the total energy, which is a factor of
$\sim 2.2-2.6$ higher. 

Notice the the above ratios of the free to the total energy are
substantially lower than those calculated by Metcalf, Leka, \& Mickey
(2005), for NOAA AR 10486, on 2003 October 29. Using the Virial
theorem and implicitly assuming NLFF magnetic fields, these
authors found that the free energy ranged from $\sim 44$\% (a few
hours before a major X10 flare) to $\sim 75-80$\% (in the course of,
and shortly after, the flare), of the total energy. These ratios
appear extraordinarily high, at least in view of eruption models 
that predict the eruption onset when the free
energy exceeds $10-15$\% of the total energy (see, for
example, DeVore \& Antiochos 2005). Of course, NOAA AR 10486 was an
extraordinary AR, which might account for its unusual behavior. 

Despite the large difference of energy and helicity budgets between
the two ARs, notice that significant magnetic helicity is 
present even in the 
noneruptive NOAA AR 8844. Indeed, the average relative helicity of
the AR is $\bar{H}_m \simeq (1.5 \pm 0.4) \times 10^{42}\;Mx^2$, 
with the helicity budget of a typical CME estimated at 
$\sim 2 \times 10^{42}\;Mx^2$ (DeVore 2000). 
With a minor helicity increase,
therefore, the AR should be capable of producing a typical CME before
relaxing to the potential state. Interestingly, a faint 
halo CME occurred
above the AR on 2000 January 26 at $\sim$ 12:00 UT and the AR started
to decay less than $24\;hr$ later, on 2000 January 27 (Schmieder et
al. 2004). As the AR was still growing during the IVM observations, it
is likely that its magnetic helicity was further increased by January
26. No significant flaring activity was associated to the CME. 

NOAA AR 9165, on the other hand, gave an eruptive M2 flare a few hours
before the IVM observations on 2000 September 15, as well as two even
stronger eruptive flares (M5.9 and M3.3) on the next day. Its relative
magnetic helicity, $\sim (-13 \pm 4) \times 10^{42}\;Mx^2$ was enough
to launch nearly seven typical CMEs. Perhaps not surprisingly, the AR
survived for several more days and could clearly be followed until it
crossed the western solar limb. 

Back in our analysis, the 
compromise brought by the LFF approximation is reflected on the
uncertainties accompanying the estimations of the free magnetic energy
in both ARs. Obviously, the free magnetic energy is a crucial
parameter in assessing the eruptive potential of a given AR (see,
e.g., Metcalf, Leka, \& Mickey 2005). With average 
linearized free energies of $(0.15 \pm 0.1) \times 10^{32}\;erg$ and 
$(1.33 \pm 0.6) \times 10^{32}\;erg$ for NOAA ARs 8844 and 9165,
respectively, the lowest uncertainties are estimated at $\sim 67$\%
and $\sim 45$\%, respectively. A generalization allowing NLFF fields
will hopefully restrict these uncertainties. 

Finally, in Figure \ref{ecomp} we compare our estimated potential and
total magnetic energies, equations (\ref{pen1}) and (35) - (36), with
those obtained by the calculation of the Virial theorem, equation
(\ref{vir}), for both ARs. To implement the Virial theorem, we perform
a current-free and a LFF extrapolation of each IVM magnetogram, the
latter using the inferred maximum-likelihood $\alpha$-value for this
magnetogram. Estimates of the Virial theorem are represented by dashed
curves and triangles. Solid curves and rectangles refer to the
linearized expressions, while dotted curves refer to the exact
expressions. For a convenient comparison, the scaling for the
potential energy (blue curves) is different than the scaling for the
total energy (red curves). From the plots in Figure \ref{ecomp}, we
first notice that our potential-energy expression, equation
(\ref{pen1}), gives almost identical results with the Virial theorem
for both ARs. The average fractional differences 
$|E_p - E_{p_{(Virial)}}|/(E_p + E_{p_{(Virial)}})$ 
are $\sim 0.7$\% and
$\sim 0.2$\% for ARs 8844 and 9165, respectively. The difference is
larger for the total energies. On average, the fractional difference
is $\sim 1.6$\% ($\sim 1.1$\%) for the linearized (exact) expressions
in NOAA AR 8844. For NOAA AR 9165, the average fractional difference
is $\sim 2.7$\% ($\sim 3.2$\%) for the linearized (exact)
expressions. These differences are small and generally within the
uncertainties in the calculation of energies. In addition, the Virial
theorem provides consistently slightly higher total energies for NOAA
AR 8844, while it consistently provides slightly lower total energies
for NOAA AR 9165. This mixed behavior, as well as the source of the
slight discrepancy in total energies, are unclear. One possible reason
may be the application of Fourier transforms, and hence an implicit
assumption of periodic boundary conditions, in analytical expressions
where fields are required to vanish at infinity. Why this does not
have an impact in the calculation of the potential energy is also
unclear. Nevertheless, the small discrepancies 
prompt us to conclude that our energy expressions are consistent with
the Virial theorem. The reasons why we have derived and used them
instead of the latter are that (i) they provide a self-consistent
description of the energy {\it and} helicity budgets, and (ii) they
are physically intuitive, derived from first principles, and,
hopefully, capable of being generalized for NLFF magnetic fields. 
\section{Summary and discussion}
The reliable calculation of the magnetic energy and helicity budgets in
the active-region solar corona is an essential step toward the
quantitative understanding of solar eruptions and has
profound space-weather applications. Our goal is to derive a practical
set of equations that are applicable to solar
vector magnetograms and can evaluate the magnetic energy and
relative magnetic helicity budgets in a physically intuitive,
self-consistent manner. Here we provide expressions 
for the magnetic energy and relative helicity
budgets in case of a constant-alpha, flux-balanced, 
magnetic structure, thus
implementing the LFF approximation. These equations are to be
generalized into magnetic structures with non-constant alpha values,
thus implementing the NLFF approximation. This objective will be
pursued in a later study. 

To perform our LFF analysis we separately derive each of the terms
present in the energy-helicity formula of Berger (1988), namely the
total magnetic energy, the potential magnetic energy, and the relative
magnetic helicity related to the free magnetic energy, together with
their uncertainties. Our analysis unifies numerous expressions for the
relative helicity and links several virtually unconnected studies 
into a self-consistent energy-helicity description that is
practical enough to be applied to vector magnetograms of solar active
regions. For the ground-state, potential, magnetic energy 
we provide a general surface-integral expression, equation
(\ref{pen1}). This expression gives results practically 
identical to those of the magnetic 
Virial theorem. The potential magnetic energy is then used
as a free parameter to explicitly 
determine the total and free magnetic energy, as
well as the relative magnetic helicity. The variable relating the
potential energy to the free energy and the relative helicity has been
calculated in two ways - an exact and a linearized one - by using and
extending the analysis of Berger (1985). As a result, the magnetic
energy and helicity budgets are calculated self-consistently as 
{\it surface integrals}, equations (35) - (37). This
development reduces significantly the required computations. 
Reliability and computational
speed are essential elements of a future real-time or near real-time 
calculation of the magnetic energy and helicity budgets 
in solar active regions. 

To test our derivations we used three different types of
semi-analytical LFF magnetic dipoles (\S4.3). The conventional  
volume-integral expression for the relative magnetic helicity was
compared with our exact and linearized surface-integral expressions.
The convincing match between the volume- and surface-integral helicity
expressions for all models implies that our formulations are
model-independent. Moreover, as 
expected from the analysis, the linearized
surface-integral expression of the relative magnetic helicity,
equation (37b), consistently provides a lower limit of the helicity
present in the magnetic structures, with this behavior being more
pronounced for large alpha values. For smaller alpha, all helicity
expressions give nearly identical results. By extension, the linearized
surface-integral expressions provide 
reasonable lower limits of the total and
free magnetic energy, equations (35b) and (36b), respectively. Given
also the convenient calculation of uncertainties in the linearized
case, equations (38b) - (40b), we conclude that linearization is
preferable over using the exact formulas, where alpha can 
resonate with the value of $(2 \pi/L)$ and hence lead to infinite
energies and helicities, as is well-known for LFF magnetic structures.

Two series of solar vector magnetograms, one for an eruptive and
another for a noneruptive active region, were thereafter subjected to
our analysis (\S5). Both the exact and linearized expressions for the
energy and helicity were used. We found that the exact expressions
tend to give more spiky temporal evolutions and hence larger
uncertainties in temporal averages, which provides 
an additional reason for preferring the linearized energy/helicity
expressions in observations. 
Both ARs happened to exhibit nearly the same absolute alpha value. The
eruptive active region, however, included several times more magnetic
flux than the noneruptive AR. This effect was greatly amplified when
the free magnetic energies and relative magnetic helicities of the two
active regions were calculated and compared. This leads us to the
conclusion that comparing the free energies and helicities might be 
a safe (safer than simply calculating the total magnetic flux) way of
distinguishing between eruptive and noneruptive active regions{\footnote{For
an alternative criterion, based on the magnetic connectivity in 
solar active regions, see Georgoulis \& Rust (2007).}}. The
crucial point, however, is the {\it reliable} calculation of
free energies and helicities. The LFF approximation is certainly not
very reliable, as can be seen from the large
error bars accompanying our free energy and helicity estimates 
($\sim 45$\% - $70$\%). To reach sound conclusions, the 
analysis involving and comparing free
magnetic energies / relative magnetic helicities must be applied to
statistically significant samples of active regions, and ideally by
utilizing the NLFF approximation. 

The force-free approximation is a prerequisite for our analysis 
because it is a very difficult, if not intractable, problem to 
perform non-force-free calculations of the nonpotential magnetic
energy and the magnetic helicity in active regions. 
The only hope for
non-force-free energy and helicity calculations emerges 
from data-driven three-dimensional
magnetohydrodynamical (3D MHD) simulations 
of the active-region corona (e.g. Abbett 2003; Roussev et
al. 2004) which, however, require immensely time-consuming
calculations. 3D MHD models are certainly capable of advancing our
physical understanding of solar eruptions but, because of their intense
computations, they cannot contribute to a real-time, or near
real-time, space weather forecasting capability. Alternatively,
non-force-free energy and helicity estimates can be obtained if an
active region is continuously observed from its formation and
thereafter. In this case, total energies and helicities can be
calculated by temporally 
integrating the Poynting flux and magnetic helicity injection
rate, respectively. If the birth of an active region is not
observed, then the initial energy and helicity can only be assumed. In 
any case, both the Poynting flux and the helicity injection rate
require the flow velocity of the magnetized plasma on the
boundary of the magnetic field measurements. Inferring a 
{\it reliable} flow velocity is a completely independent, as well as
highly nontrivial, problem (for a review, see Welsch et al. 2007). 

Our force-free equations are
physically better suited to apply to chromospheric, rather than
photospheric, vector magnetograms. It has yet to be established
whether the NLFF approximation holds for the active-region
chromosphere (see Metcalf et al. [1995] in conjunction with
Socas-Navarro [2005]) but it is almost certainly more valid there than in
the photosphere. The first high-quality chromospheric vector
magnetograms have already been obtained (the above authors as well as 
Leka \& Metcalf 2003; Metcalf, Leka, \& Mickey 2005; Wheatland \&
Metcalf 2006) but a routine acquisition and reduction 
of such data may still be a task for the future. 
In brief, force-free equations may be applied to 
photospheric vector magnetograms as a zero-order (LFF) or first-order
(NLFF) approximation, but one expects larger uncertainties in the
values of energy and helicity budgets, 
than when chromospheric vector magnetograms are used. 

Concluding, we emphasize that the present analysis cannot 
fully uncover the importance of magnetic helicity in solar eruptions. 
Here we only show two examples that appear to
point to this direction but an answer would require large numbers of
active regions and NLFF energy/helicity equations, as already said. 
Our objective here was to calculate the relative magnetic
helicity in active regions as an integral part of the energetics and
complexity of the studied magnetic structures. It would 
be an important leap forward if it was convincingly shown 
that flare- and CME-prolific active regions 
exhibit significant quantitative
differences in their free magnetic energy and/or total relative
helicity (large free magnetic energy does not necessarily imply a
large total relative magnetic helicity because roughly equal and opposite
amounts of helicity may be simultaneously present - see Phillips,
MacNeice, \& Antiochos [2005]) compared to quiescent ARs.
Intriguing clues to this direction stem from the study of the
structural magnetic complexity in solar active regions
(e.g. Georgoulis 2005b; Abramenko 2005) or the
calculation of the free magnetic energy in active regions with
exceptional flare and CME records (Metcalf, Leka, \& Mickey 2005) but
the role of helicity is yet to be uncovered. Some pieces of
evidence suggesting the importance of helicity in solar eruptions 
stem from the frequent presence of sigmoids in eruptive active regions 
(Rust \& Kumar 1996; Canfield, Hudson, \& McKenzie 1999), 
apparently due to significant amounts of helicity with a 
{\it prevailing} sign, the presence of large and highly variable alpha
values in eruptive active regions (Nindos \& Andrews
2004), and the statistical correlation between large
helicity injection rates and X-class flares/CMEs 
(LaBonte, Georgoulis, \& Rust 2007). 
Our forthcoming NLFF analysis will be well suited to
address the role of helicity in solar eruptions and we intend to 
carry out this study in the future. 
\acknowledgements
This work is dedicated to the memory of its co-author, Barry
J. LaBonte. Barry is remembered as a deeply knowledgeable, distinguished
colleague and an inspiring mentor. I am grateful to D. M. Rust for our 
continuous interaction on magnetic helicity in the Sun and for a
critical reading of the manuscript. I also thank 
A. Nindos and S. R\'{e}gnier for clarifying discussions on helicity
issues and an anonymous referee whose numerous critical comments and
suggestions resulted in substantial improvements in the paper. 
Partial support for this work has been received by NASA 
Grants NAG5-13504 and NNG05-GM47G.
\appendix
\section{Equivalence of equations (\ref{ehf1}) and (\ref{ehf}) for 
the energy-helicity formula in the linear force-free approximation} 
To show that equations (\ref{ehf1}) and (\ref{ehf}) are equivalent in
the LFF approximation, it is sufficient to show that the potential
energy $E_p$ is given by 
\beq
E_p={{1} \over {8 \pi}} 
\int _{\cs} \mbf{B} \times \mbf{A_p} \cdot \mbf{\hat{z}} \;d \cs\;\;,
\label{ap1}
\eeq
for any LFF magnetic field $\mbf{B} \ne \mbf{B_p}$. 

We first decompose $\mbf{B}$ in equation (\ref{ap1}) into its
potential (poloidal) and nonpotential (toroidal) components, 
$\mbf{B_p}$ and $\mbf{B_c}$. Then, equation (\ref{ap1}) becomes 
\beq
E_p={{1} \over {8 \pi}} 
\int _{\cs} \mbf{B_p} \times \mbf{A_p} \cdot \mbf{\hat{z}} \;d \cs + 
{{1} \over {8 \pi}}
\int _{\cs} \mbf{B_c} \times \mbf{A_p} \cdot \mbf{\hat{z}} \;d \cs\;\;.
\label{ap2}
\eeq
The first integral of equation (\ref{ap2}) is already
the potential energy as shown in equation (\ref{pen1}). To prove
equation (\ref{ap1}), therefore, it is sufficient to show that 
\beq
\int _{\cs} \mbf{B_c} \times \mbf{A_p} \cdot \mbf{\hat{z}} \;d \cs =0\;\;.
\label{ap3}
\eeq

From $\mbf{B_p} \cdot \mbf{B_c}=0$, we construct the volume integral 
$\int _{\cv} \mbf{B_p} \cdot \mbf{B_c} d \cv =0$. Substituting the
definition of $\mbf{A_p}$ from equation (5a) into this volume
integral, we find after some analysis that 
\beq
\int _{\partial \cv} \mbf{A_p} \times \mbf{B_c} \cdot \mbf{\hat{n}} d \sigma = 
- \int _{\cv} \mbf{A_p} \cdot \nabla \times \mbf{B_c} d \cv\;\;.
\label{ap4}
\eeq
Taking into account that (i) $\mbf{A_p}$ vanishes at infinity, and
(ii) $\nabla \times \mbf{B_c} = \nabla \times \mbf{B}$, because 
$\nabla \times \mbf{B_p} = 0$, equation (\ref{ap4}) further reduces to 
\beq
\int _{\cs} \mbf{B_c} \times \mbf{A_p} \cdot \mbf{\hat{z}} d \cs = 
- \int _{\cv} \mbf{A_p} \cdot \nabla \times \mbf{B} d \cv\;\;.
\label{ap5}
\eeq
In the LFF approximation, however, 
$\nabla \times \mbf{B} = \alpha \mbf{B}$, with $\alpha$ constant, so
equation (\ref{ap5}) gives 
\beq
\int _{\cs} \mbf{B_c} \times \mbf{A_p} \cdot \mbf{\hat{z}} d \cs = 
- \alpha \int _{\cv} \mbf{A_p} \cdot \mbf{B} d \cv\;\;.
\label{ap6}
\eeq
Given the gauge-invariant definition of the relative magnetic
helicity, however, it can be shown (Berger 1988; 1999) that 
\beq
\int _{\cv} \mbf{A_p} \cdot \mbf{B} d \cv=0\;\;.
\label{ap7}
\eeq
Combining equations (\ref{ap6}) and (\ref{ap7}) we obtain equation
(\ref{ap3}). Therefore, equation (\ref{ap1}) is true and hence
equations (\ref{ehf1}) and (\ref{ehf}) in \S3 are equivalent as
asserted. 
\section{Derivation of the variable linking the potential and the
  total magnetic energy in the LFF approximation}
Here we will derive the form of the dimensionless variable $f$ in
equation (\ref{es1}). This variable links the total and the free 
magnetic energies in a constant-alpha magnetic structure.
We will use and extend
the analysis performed in Appendix AII of Berger (1985). 
Assuming planar geometry, Berger (1985) utilized 
Chandrasekhar's (1956; 1961) decomposition of an arbitrary magnetic
field vector into a poloidal and a toroidal components 
and, in view of the LFF approximation, he derived  
the total magnetic energy and the relative magnetic 
helicity. 

Following Berger (1985), the 
total magnetic energy of the structure is given by 
\beq
E = {{\pi} \over {2}} \sum _{l=1} ^{n_x} \sum _{m=1} ^{n_y} 
{{|b^2_{u_l,v_m}|} \over {k_{l,m}}}\;\;,
\label{apc1}
\eeq
where $b_{u_l,v_m}$ is the Fourier amplitude of the vertical
magnetic field $B_z$ for the harmonic $(u_l,v_m)$ in a two-dimensional
Fourier space with linear dimensions $n_x$, $n_y$. In addition, we
have $k^2_{l,m}=u^2_l+v^2_m - \alpha '^2$. The force-free
parameter $\alpha '$ is expressed in inverse length units  
(i.e., $1/x$, where $x$ is the number of unit lengths required for
$\alpha '=1$) and not in physical units. This is why it is represented
by $\alpha '$, while the $\alpha$ used so far 
refers to the force-free parameter
expressed in physical units. Typically, $\alpha ' = \alpha d$, where
$d$ is the unit length expressed in physical units. 
Berger (1985) assumes periodic boundary conditions and a 
length unit of $[L/(2 \pi)]$, where $L$ 
is the linear dimension of the magnetic
structure on the boundary $\cs$. Moreover, $u_l=(2 \pi l /L)$ 
and $v_m=(2 \pi m /L)$. Then, the direct and inverse Fourier transform
of $B_z$ can be performed on $\cs$ only so that one can write 
\beq
B_z(x,y) |_{\cs} = \sum _{l=1} ^{n_x} \sum _{m=1} ^{n_y}
b_{u_l,v_m}e^{i(u_lx+v_my)}\;\;.
\label{apc2}
\eeq
The required boundary conditions for $b_{u_l,v_m}$ in order to have a
real and finite magnetic energy and helicity is $b_{u_l,v_m}=0$ for 
$\sqrt{u^2 + v^2} \le |\alpha '|$ (see also Alissandrakis 1981). 

Assuming that the magnetic structure does not include electric
currents ($\alpha '=0$), then equation (\ref{apc1}) provides the
potential magnetic energy of the structure, namely 
\beq
E_p = {{\pi} \over {2}} \sum _{l=1} ^{n_x} \sum _{m=1} ^{n_y} 
{{|b^2_{u_l,v_m}|} \over {q_{l,m}}}\;\;,
\label{apc3}
\eeq
where $q^2_{l,m}=u^2_l+v^2_m$. 

From equation (\ref{es1}), the variable $f$ is given by the
dimensionless ratio 
\beq
f = {{E - E_p} \over {E_p}}\;\;.
\label{apc4}
\eeq
Substituting equations (\ref{apc1}) and (\ref{apc3}) into equation
(\ref{apc4}) we obtain 
\beq
f= {{\sum _u \sum _v |b^2_{u,v}| {{q-k} \over {kq}}} \over 
    {\sum _u \sum _v {{|b^2_{u,v}|} \over {q}}}}\;\;, 
\label{apc5}
\eeq
where we have denoted $\sum _{l=1}^{n_x} \sum _{m=1}^{n_y}$ by 
$\sum _u \sum _v$ for simplicity. The ratio of sums in equation
(\ref{apc5}) depends on $\alpha '$ because of its dependence on
$k$. This dependence can cause problems when 
$|\alpha '| \rightarrow (2 \pi/L)$ because 
$k \rightarrow 0$ for $l=m=1$ in this case and $f$ becomes infinite. 
This problem is not new; that LFF fields sometimes give solutions that
are not fully specified by the boundary condition and may include
infinite energy has been explicitly acknowledged by Alissandrakis
(1981), but also by Chiu \& Hilton (1977), using a different
analytical framework. Clearly, this is a caveat of the LFF
approximation and restricts its applicability. 
To avoid infinite energy values when 
$|\alpha '| \rightarrow (2 \pi/L)$, equation (\ref{apc5}) can 
be linearized with respect to ${\alpha '}^2$, 
assuming small values of $\alpha '$. We first write 
$(q-k)/(kq)=(q^2-kq)/(kq^2)$. Expanding $q^2 -kq$ in a MacLaurin
series, one finds $q^2 - kq \simeq (q^2 - k^2)/2 =\alpha '^2 /2$. Moreover,
for $|\alpha '| \ll \sqrt{2} 2 \pi /L$, one finds $k q^2 \simeq q^3$. 
Then, the linearized equation (\ref{apc5}) becomes 
\beq
f_l= {{\alpha '^2} \over {2}} {{\sum _u \sum _v {{|b^2_{u,v}|} \over {q^3}} } \over 
                            {\sum _u \sum _v {{|b^2_{u,v}|} \over {q}}}}\;\;. 
\label{apc7}
\eeq

Berger (1985) goes further on to derive a linearized 
expression for the total
relative magnetic helicity in the volume $\cv$ above $\cs$, namely
\beq
H_m = 4 \pi ^2 \alpha ' \sum _u \sum _v {{|b^2_{u,v}|} \over {kq^2}}\;\;.
\label{apc8}
\eeq
Although not explicitly mentioned in Berger's (1985) 
analysis, equation (\ref{apc8}) appears to occur by assuming that 
$q^2 - kq \simeq \alpha '^2$, instead of 
$q^2 - kq \simeq \alpha '^2/2$, that we have assumed
in equation (\ref{apc7}). In our formulation, therefore, 
Berger's (1985) equation (\ref{apc8}) is a factor of two too
high. If no linearization is performed, then our equation (\ref{apc5})
is in agreement with Berger's (1985) analysis. 

In an observed vector magnetogram, the best $\alpha$-value is inferred
in physical units of inverse length. The scaled value $\alpha '$ of
the force-free parameter relates to $\alpha$ via the equation 
$\alpha ' = \alpha d$ where $d$ is the elementary length in the
magnetogram, expressed in physical units. As in any discrete 
parameter distribution with a well-defined (preferably fixed) 
length element, the length $d$ in the magnetogram
can be naturally represented by the linear size of the magnetogram's
pixel. From this understanding and using the definition $q^2=u^2+v^2$, 
the linearized expression $f_l$ for $f$ can be written as 
\beq
f_l= \mathcal{F}_l d^2 \alpha ^2\;\;\;where\;\;\;\;\;
\mathcal{F}_l= {{1} \over {2}}
{{\sum _u \sum _v {{|b_{u,v}^2|} \over {(u^2 + v^2)^{3/2}}}}
\over {\sum _u \sum _v {{|b_{u,v}^2|} \over {(u^2 + v^2)^{1/2}}}}}\;\;.
\label{apc9}
\eeq

It is important to emphasize that the 
linearization $f_l$ as shown in equation (\ref{apc9}) 
provides a lower limit of $f$, and
hence a lower limit of the free magnetic energy $E_c$ and the relative
magnetic helicity $H_m$ in the LFF approximation (equations (35) - (37)).
Green et al. (2002) and D\'{e}moulin et al. (2002) reached the same
conclusion when deriving the linearized helicity expression of
equation (\ref{in2}). Since the LFF magnetic energy is the minimum energy
for a given relative helicity (a consequence of the Woltjer-Taylor
theorem), the linearized energy/helicity expressions are
underestimations of the actual energy/helicity values. 
The underestimation of $E_c$ and $H_m$ is
negligible for small values of $|\alpha '|$ and increases as 
$|\alpha '| \rightarrow (2 \pi /L)$. This, however, does not
invalidate the linearized energy/helicity expressions for larger
$|\alpha '|$. As we see in Figure \ref{hc}b, the underestimation
factor is reasonable even for large $|\alpha '|$, at least in view of
other sources of uncertainties that are expected for observed magnetic
configurations, and especially the use of the LFF approximation
itself. D\'{e}moulin (2006), 
also provides a practical explanation of the underestimation
effect based on well-known properties of the LFF magnetic fields.   

\clearpage
\newpage
\centerline{\includegraphics[width=11.5cm,height=13.5cm,angle=0]{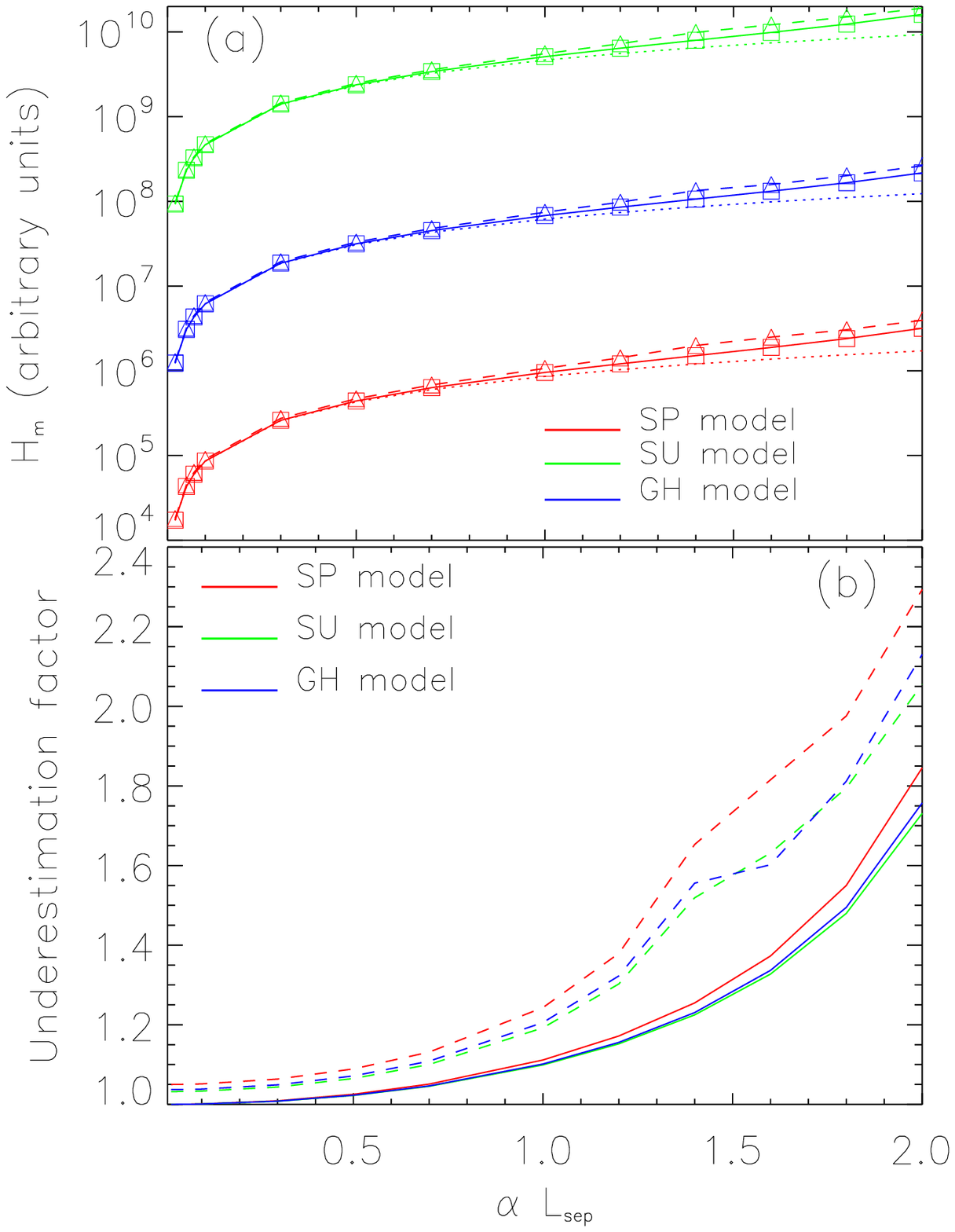}}
\figcaption{Comparison between the volume- and the surface-integral 
expressions of the relative magnetic helicity $H_m$ in three LFF
dipole models with analytical boundary conditions for the vertical
magnetic field component. Boundary conditions are taken by 
the submerged poles (SP) model solution, the Sakurai-Uchida 
(SU) solenoidal model solution, and the Gold-Hoyle (GH) model 
solution (see \S4.3 for details). 
Different model results are represented by different colors.
(a) Dashed curves and
triangles refer to results of the volume-integral helicity, 
equation (\ref{hmv}), while the 
exact surface-integral results, equation (40a), 
are shown by solid curves and rectangles. 
The linearized surface-integral results, equation (40b), are shown by
dotted curves. (b) Dashed curves refer to the ratios between the
volume-integral helicity and the linearized surface-integral
helicity, while solid curves show the ratios between the exact 
surface-integral helicity and the linearized surface-integral
helicity for each of the models. In
all cases, the footpoint separation length $L_{sep}$ of the dipoles has been
kept fixed. 
\label{hc}}
\newpage
\centerline{\includegraphics[width=18.cm,height=7.5cm,angle=0]{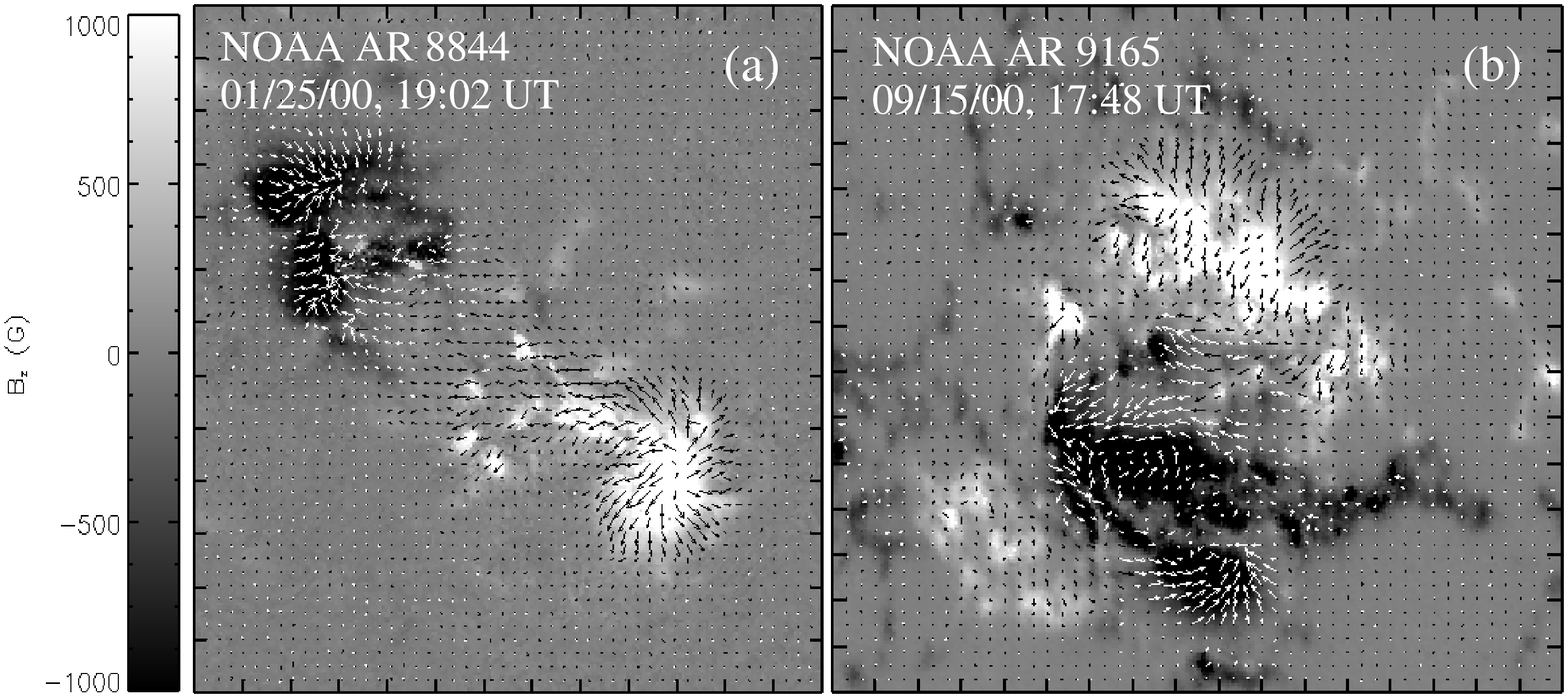}}
\figcaption{Parts of the magnetic configuration of the two studied solar
  ARs. Shown are the heliographic magnetic field components of the ARs
  on the heliographic plane. (a) Disambiguated photospheric vector
  magnetogram of NOAA AR 8844 as obtained by the IVM on 2000 January 25 at
  19:02 UT. A vector length equal to the tick mark
  separation corresponds to a horizontal magnetic field of $2300\;G$. 
  (b) Disambiguated photospheric vector
  magnetogram of NOAA AR 9165 as obtained by the IVM on 2000 September
  15 at 17:48 UT. A vector length equal to the tick mark
  separation corresponds to a horizontal magnetic field of $1760\;G$. 
  Tic mark separation in both images is $10\arcsec$. North is up; west
  is to the right. 
\label{8844}}
\newpage
\centerline{\includegraphics[width=11.cm,height=7.5cm,angle=0]{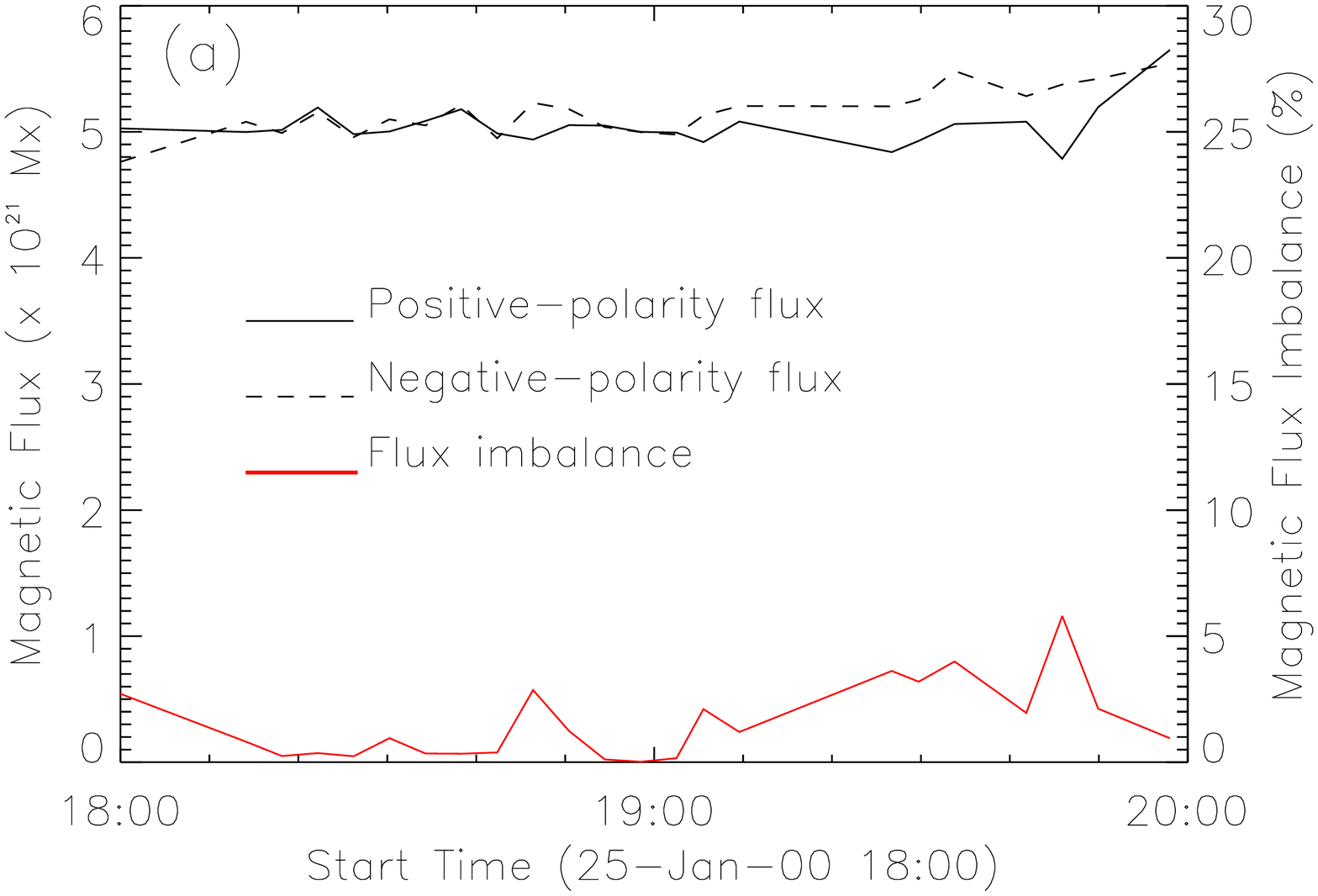}}
\centerline{\includegraphics[width=11.cm,height=7.5cm,angle=0]{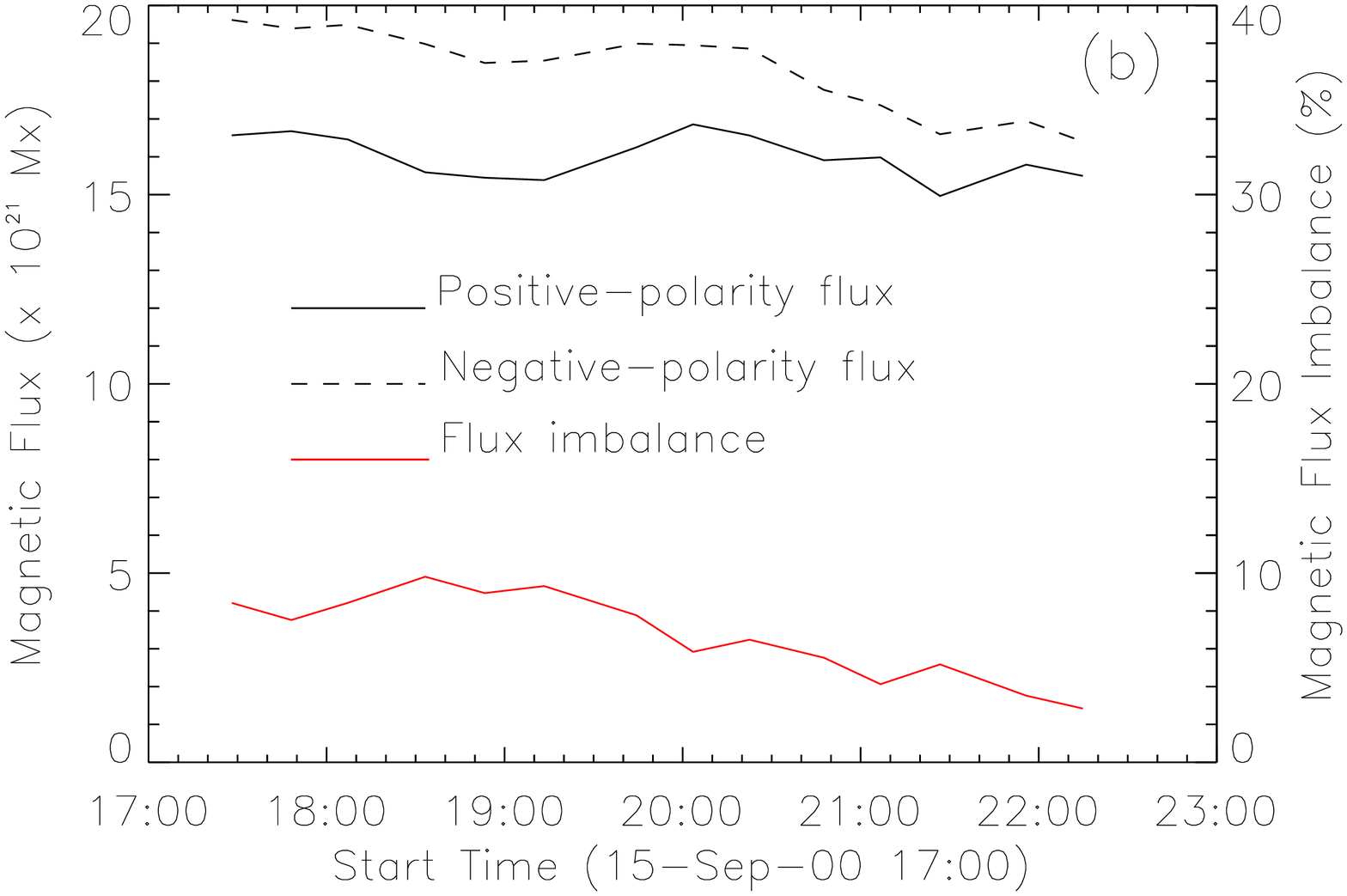}}
\figcaption{Timeseries of the magnetic flux and flux imbalance in the
  two studied ARs. Negative- (positive-) polarity fluxes are indicated
  by dashed (solid) curves, with readings on the left ordinate, 
  while the red line corresponds to
  the relative magnetic flux imbalance in the ARs, with readings on the
  right ordinate. (a) NOAA AR 8844 (b) NOAA AR 9165.
\label{flen_fig}}
\newpage
\centerline{\includegraphics[width=12.cm,height=16.cm,angle=0]{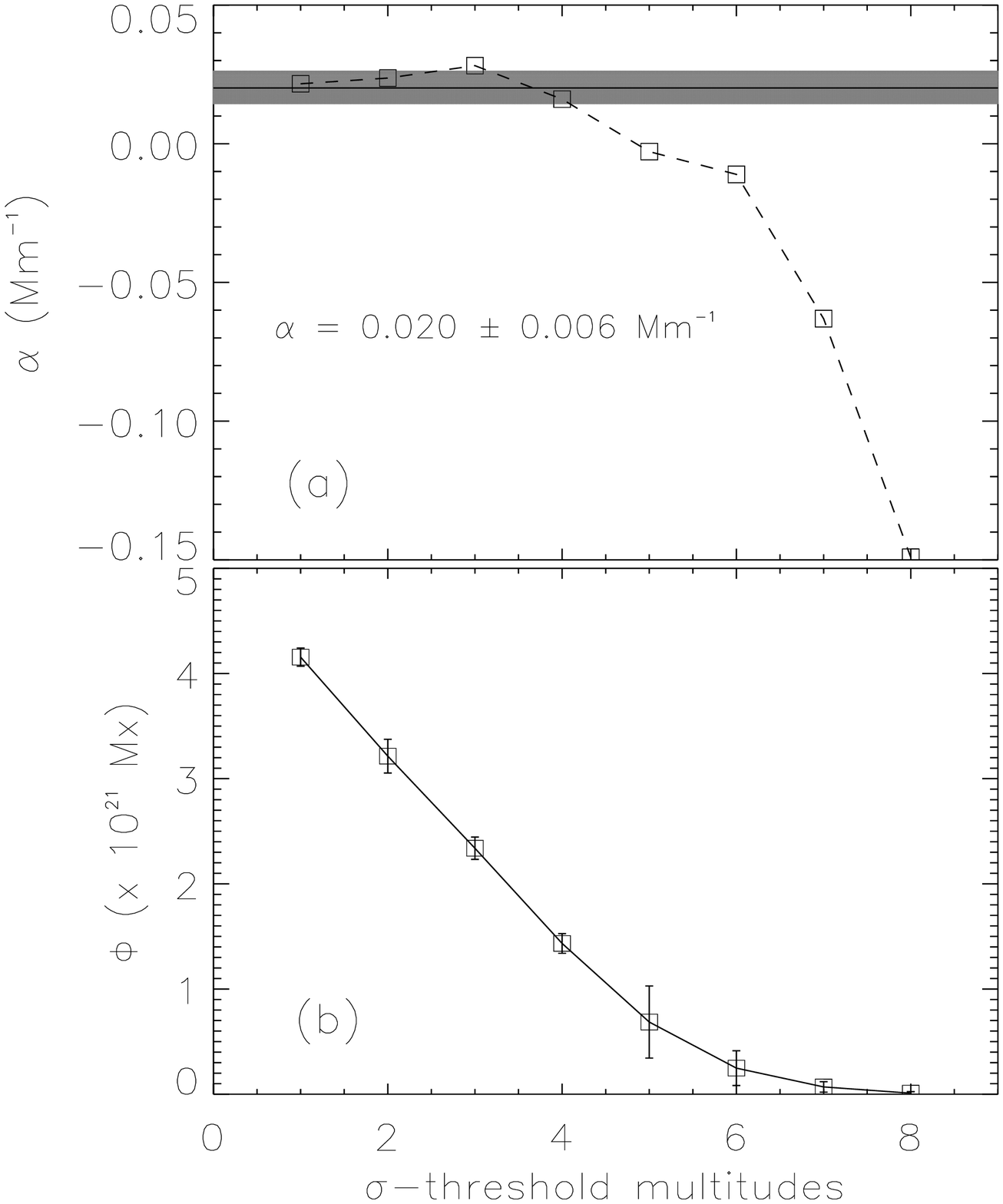}}
\figcaption{Example calculation of a unique $\alpha$-value in NOAA
 AR 8844 for a vector magnetogram obtained at 18:18 UT on 2000 January
 25. (a) Various $\alpha$-values obtained for various significance thresholds
 $\sigma$ (see text). The straight solid line indicates the
 flux-weighted average of these $\alpha$-values and the shaded area
 indicates the uncertainty in the calculation of this average. 
 (b) Estimates of the total unsigned magnetic flux in the AR for various 
 significance thresholds $\sigma$. A threshold of $1 \sigma$ corresponds to a
 vertical magnetic field of $100\;G$ and a horizontal magnetic field
 of $200\;G$. 
\label{acalc}}
\newpage
\centerline{\includegraphics[width=10.5cm,height=8.5cm,angle=0]{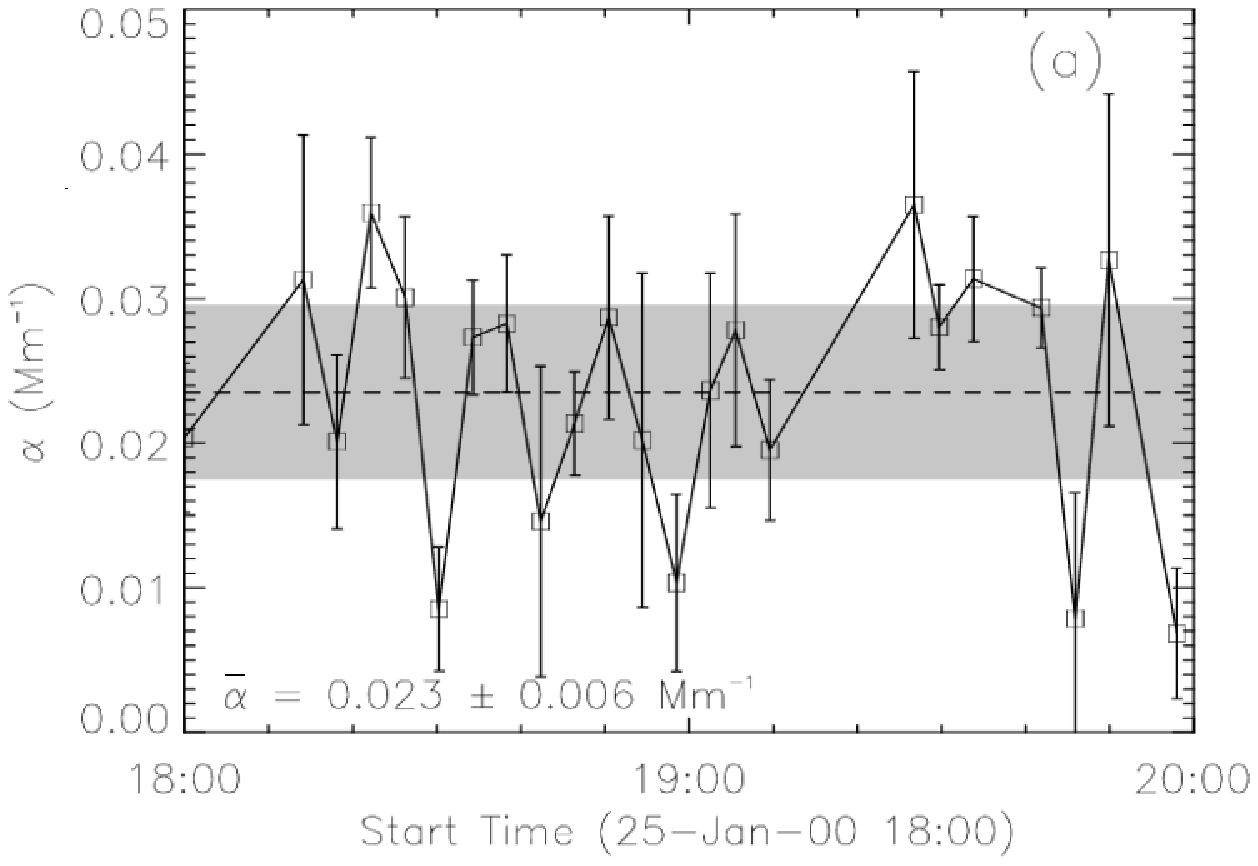}}
\centerline{\includegraphics[width=11.cm,height=8.5cm,angle=0]{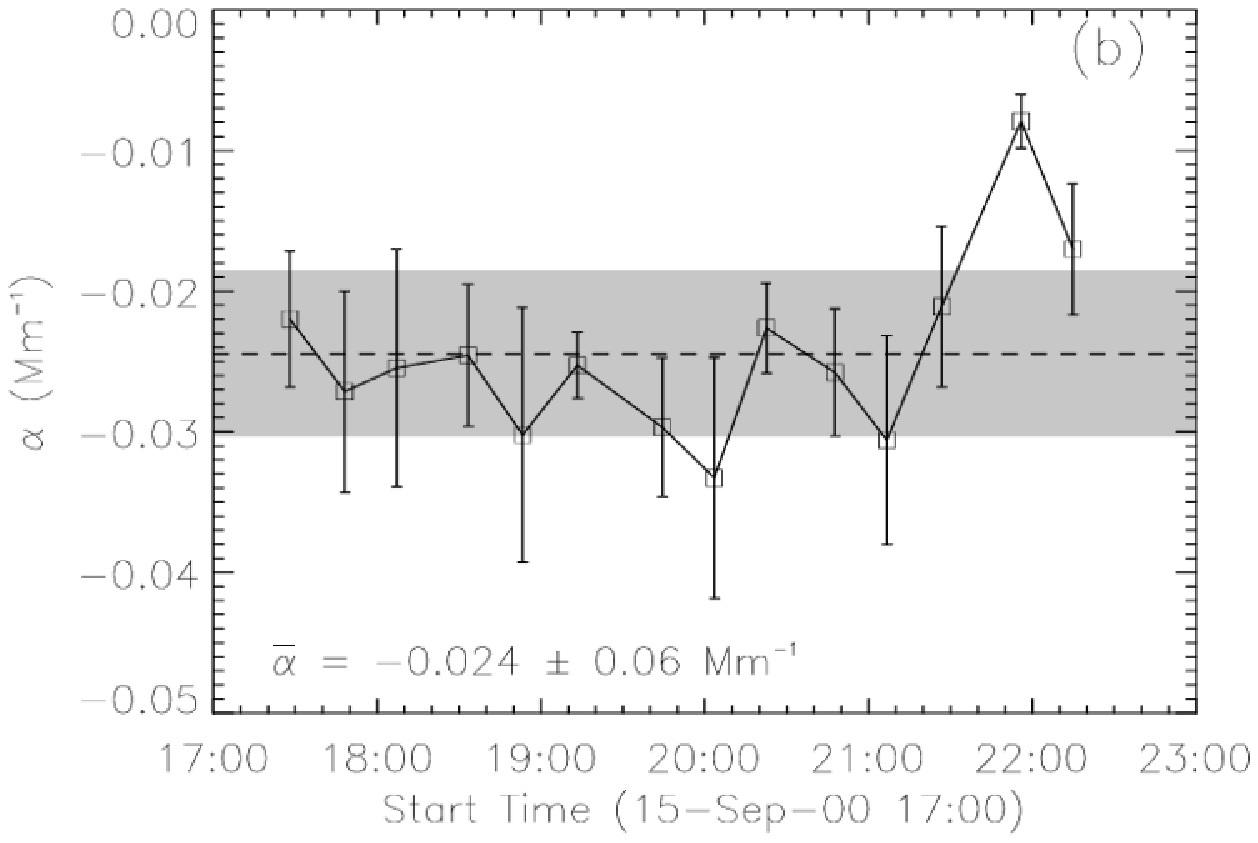}}
\figcaption{Timeseries of the constant force-free parameter $\alpha$ in
the two studied ARs. The dashed line and the surrounding shaded area
correspond to the estimated average $\alpha$-value and its
uncertainties, respectively. (a) NOAA AR 8844 (b) NOAA AR 9165.
\label{alp}}
\newpage
\centerline{\includegraphics[width=11.cm,height=15.cm,angle=0]{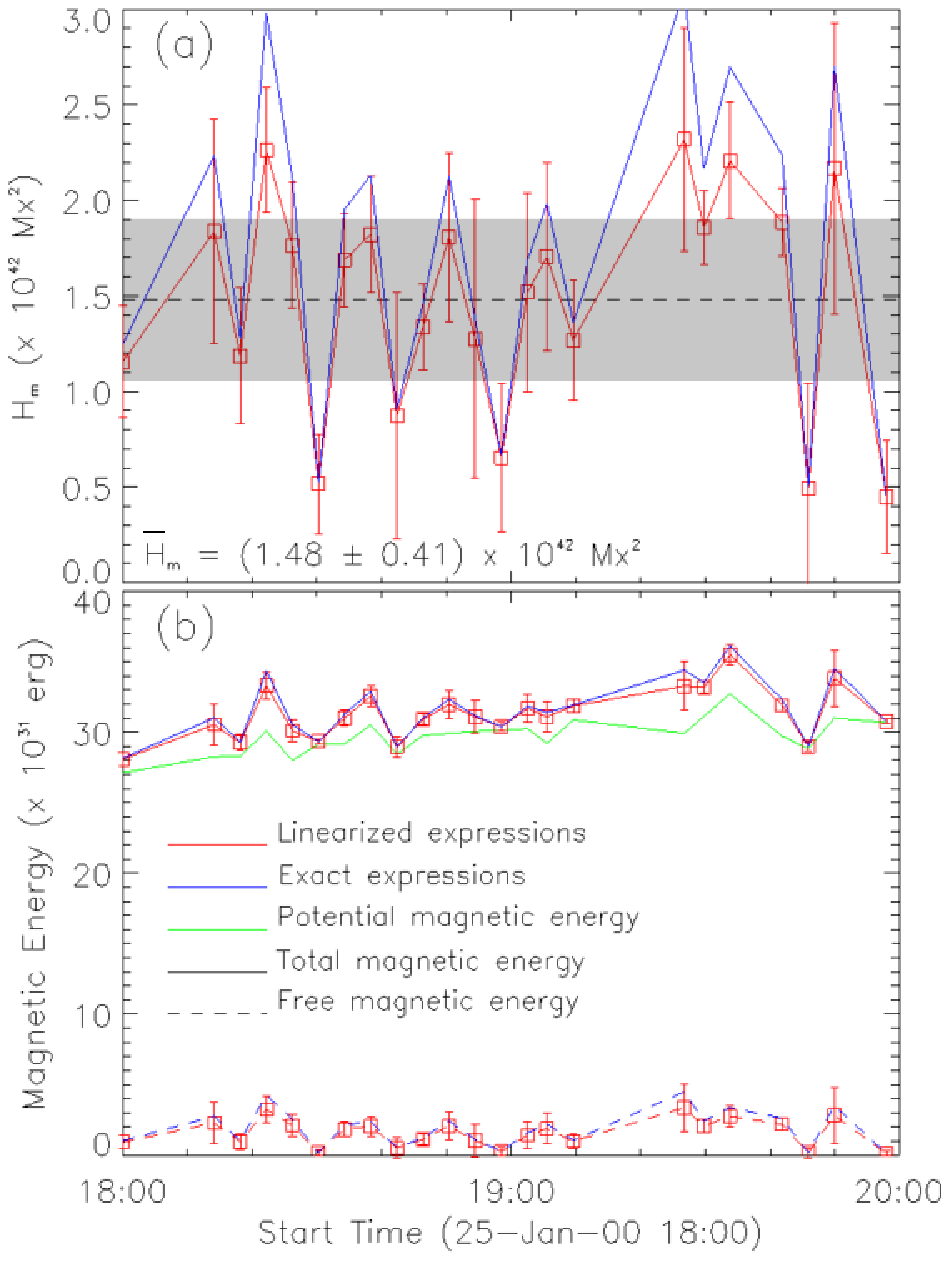}}
\figcaption{Magnetic energy and helicity budgets in
  NOAA AR 8844. Red (blue) curves correspond to the linearized
  (exact) surface-integral expressions. The error bars have been
  calculated from the linearized expressions.  
  (a) Timeseries of the total relative magnetic helicity $H_m$. The dashed 
  line and the surrounding shaded area correspond to the linearized average
  value and its uncertainties, respectively. (b) Timeseries of the
  magnetic energy budgets in the AR. The potential magnetic energy is
  shown by the green curve. The total (free) energy and its uncertainties are
  shown by the solid (dashed) curves. 
\label{hen1}}
\newpage
\centerline{\includegraphics[width=11.cm,height=15.cm,angle=0]{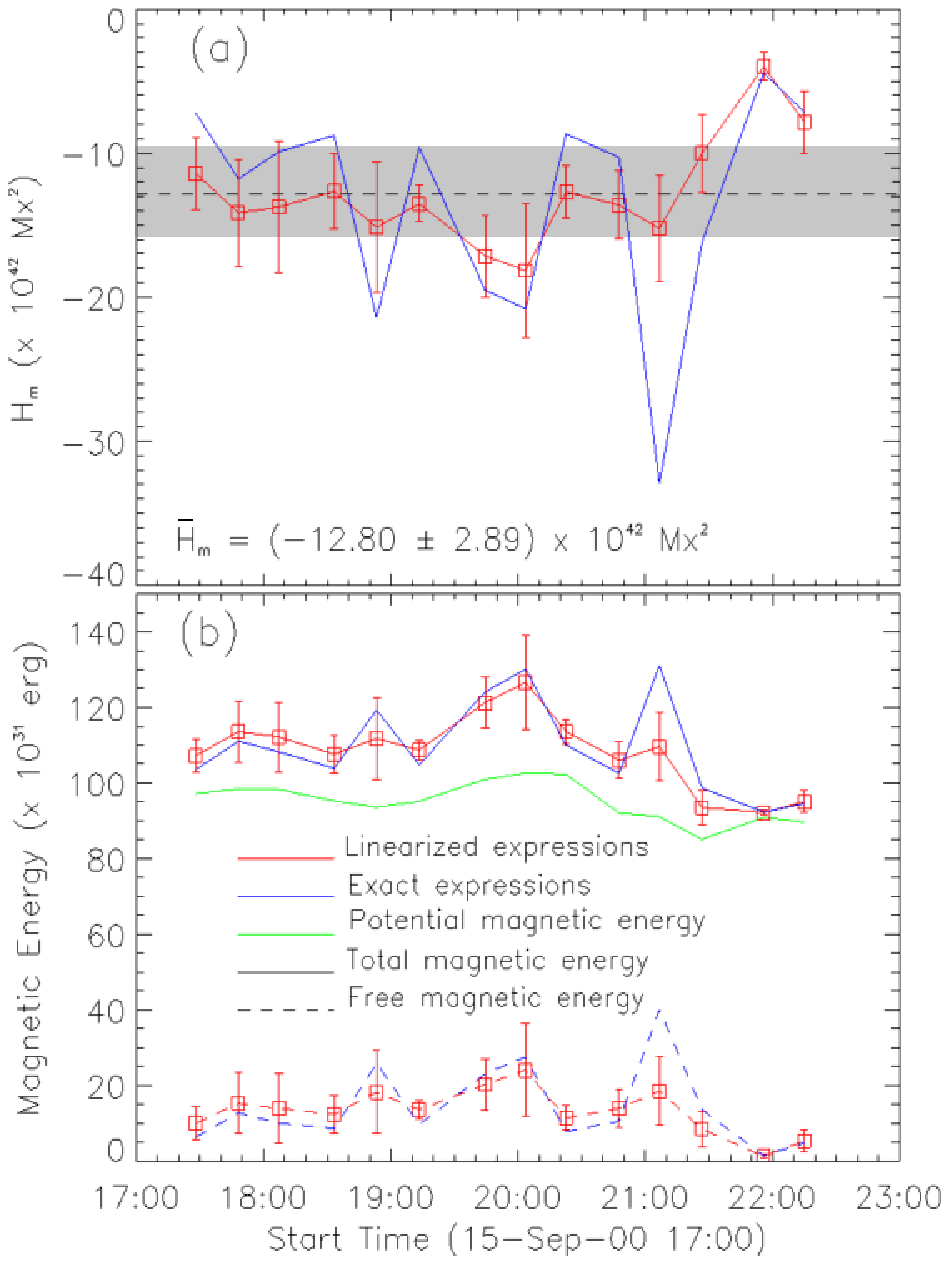}}
\figcaption{Same as Figure \ref{hen1}, but for NOAA AR 9165.
\label{hen2}}
\newpage
\centerline{\includegraphics[width=11.cm,height=7.5cm,angle=0]{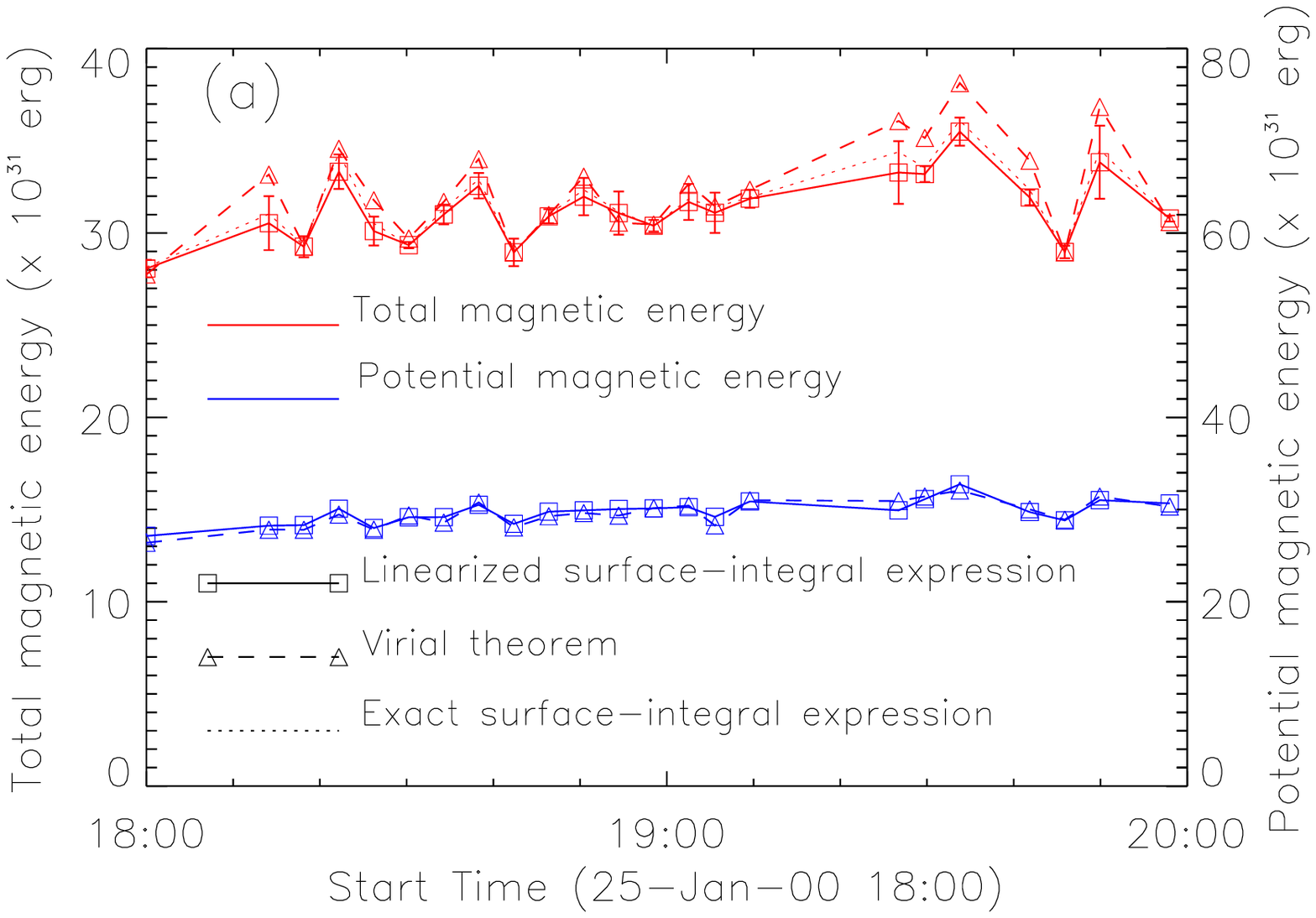}}
\centerline{\includegraphics[width=11.cm,height=7.5cm,angle=0]{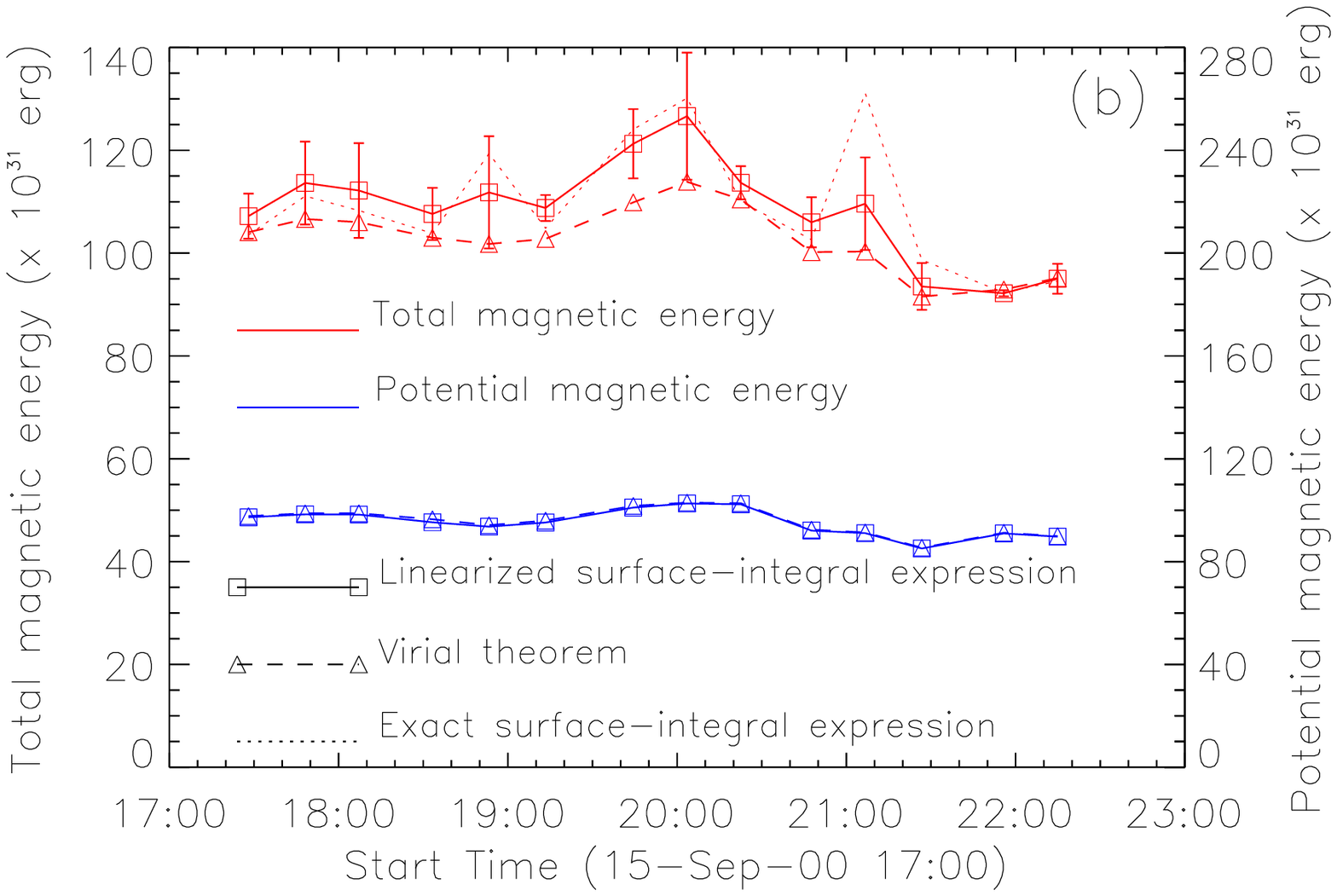}}
\figcaption{Comparison between the potential and the total magnetic energy
  estimates provided by our analysis (equations (\ref{pen1}) and
  (38a), (38b)) and the Virial theorem (equation (\ref{vir}) for the
  two studied ARs. Shown with blue (red) curves are the potential
  (total) magnetic energy estimates. The linearized (exact)
  estimates for the total energy are represented by solid curves and
  rectangles (dotted curves). The Virial-theorem estimates are
  represented by dashed curves and triangles. The error bars
  correspond to the linearized expression for the total energy. 
  For clarity in comparing the different energy values, we have applied
  a different scaling for the total energy (with readings on the left
  ordinate) than the scaling for the potential energy (with readings
  on the right ordinate). (a) NOAA AR 8844. (b) NOAA AR 9165. 
\label{ecomp}}
\newpage
\newpage
\begin{table}
\begin{tabular}{lcccc}
\hline
\hline
NOAA AR & $\bar{\Phi}$ ($\times 10^{21}\;Mx$) & $\bar{\alpha}$ ($Mm^{-1}$) &
\multicolumn{2}{c}{$\bar{H}_m$ ($\times 10^{42}\;Mx^2$)} \\
 & & & Exact & Linearized \\
\hline
8844...... & $5.1 \pm 0.2$ & $\;\;0.023 \pm 0.006$ & $\;\;\;\;1.73 \pm 0.8$ & $\;\;\;\;1.48 \pm 0.4$ \\
9165...... & $17.1 \pm 0.8$ & $-0.024 \pm 0.006$ & $-13.5 \pm 7.8$ & $-12.8 \pm 3.7$\\
\hline
Ratio..... & $\mbf{3.4 \pm 0.06}$ & $\mbf{1}$ & $\mbf{7.8 \pm 0.7}$ & $\mbf{8.6 \pm 0.5}$\\
\hline
\hline
\end{tabular}
\caption{Synopsis of the average magnetic flux, $\alpha$-value, and relative
  magnetic helicity budgets for NOAA ARs 8844 and 9165. The third row
  refers to the ratio $|P_{9165}/P_{8844}|$ between a given parameter
  $P_{9165}$ of NOAA AR 9165 and the respective parameter $P_{8844}$
  of NOAA AR 8844.}
\label{Tb1}
\end{table} 
\begin{table}
\begin{tabular}{lccccc}
\hline
\hline
NOAA AR & $E_p$ ($\times 10^{32}\;erg$) & 
\multicolumn{2}{c}{$\bar{E}_c$ ($\times 10^{32}\;erg$)} &
\multicolumn{2}{c}{$\bar{E}$ ($\times 10^{32}\;erg$)} \\
 & & Exact & Linearized & Exact & Linearized\\
\hline
8844...... & $2.97 \pm 0.1$ & $0.19 \pm 0.13$ & $0.15 \pm 0.1$ & 
$3.16 \pm 0.2$ & $3.13 \pm 0.2$\\ 
9165...... & $9.52 \pm 0.5$ & $1.44 \pm 1.1$ & $1.33 \pm 0.6$ & 
$10.96 \pm 1.2$ & $10.85 \pm 0.9$\\ 
\hline
Ratio..... & $\mbf{3.2 \pm 0.06}$ & $\mbf{7.6 \pm 1}$ & $\mbf{8.9 \pm 0.8}$ &
$\mbf{3.5 \pm 0.2}$ & $\mbf{3.5 \pm 0.1}$ \\
\hline
\hline
\end{tabular}
\caption{Synopsis of the average potential, free, and total magnetic
  energy budgets, respectively, for NOAA ARs 8844 and 9165. The third row
  refers to the ratio $(P_{9165}/P_{8844})$ between a given parameter
  $P_{9165}$ of NOAA AR 9165 and the respective parameter $P_{8844}$
  of NOAA AR 8844.} 
\label{Tb2}
\end{table}
\end{document}